\documentclass[twocolumn,10pt,aps,prb,floatfix,groupedaddress,superscriptaddress]{revtex4-2}
\usepackage{tikz,bm,bbm,bbold,graphicx,amssymb,amsmath,amsfonts,dsfont,color,braket}
\usepackage[colorlinks,allcolors=blue,breaklinks=true]{hyperref}

\DeclareMathOperator{\Tr}{Tr}

\renewcommand{\L}{\mathcal{L}}

\hypersetup{allcolors=blue,colorlinks}
\begin{document}

\newcommand{\unilu}{\affiliation{Department of Physics and Materials Science,
University of Luxembourg, L-1511 Luxembourg, Luxembourg}}
\newcommand{\UBA}{\affiliation{Universidad de Buenos Aires, Facultad de Ciencias Exactas y Naturales, Departamento de Física. Buenos Aires, Argentina.}
\affiliation{CONICET - Universidad de Buenos Aires, Instituto de Física de Buenos Aires (IFIBA). Buenos Aires, Argentina.}}

\author{Sadeq S. Kadijani}
\unilu
\author{Nicolás Del Grosso}
\UBA
\author{Thomas L. Schmidt}
\unilu
\author{M. Belén Farias}
\unilu

\title{Dynamical Casimir cooling in circuit QED systems}

\begin{abstract}
 A transmission line coupled to an externally driven superconducting quantum interference device (SQUID) can exhibit the Dynamical Casimir Effect (DCE). Employing this setup, we quantize the SQUID degrees of freedom and show that it gives rise to a three-body interaction Hamiltonian with the cavity modes. By considering only two interacting modes from the cavities we show that the device can function as an autonomous cooler where the SQUID can be used as a work source to cool down the cavity modes. Moreover, this setup allows for coupling to all modes existing inside the cavities, and we show that by adding two other extra modes to the interaction with the SQUID the cooling effect can be enhanced.     
\end{abstract}
\maketitle

\section{Introduction}

Among the different platforms that allow for an experimental study of quantum phenomena, one that has attracted much interest over the past years is quantum electrical circuits \cite{devoret1995quantum, devoret_circuit_2014,devoret2004superconducting,beaudoin2011dissipation,majer2007coupling}. These systems, which consist of quantized lumped elements such as inductors and capacitors, exhibit a wide range of applications, ranging from performing quantum information and computation tasks \cite{wustmann_parametric_2013, fosco_vacuum_2013} to the field of quantum thermodynamics \cite{senior2020heat,karimi2016otto,kadijani2020heat}.

\begin{figure*}[t!]
    \centering
    \includegraphics[width=.8\linewidth]{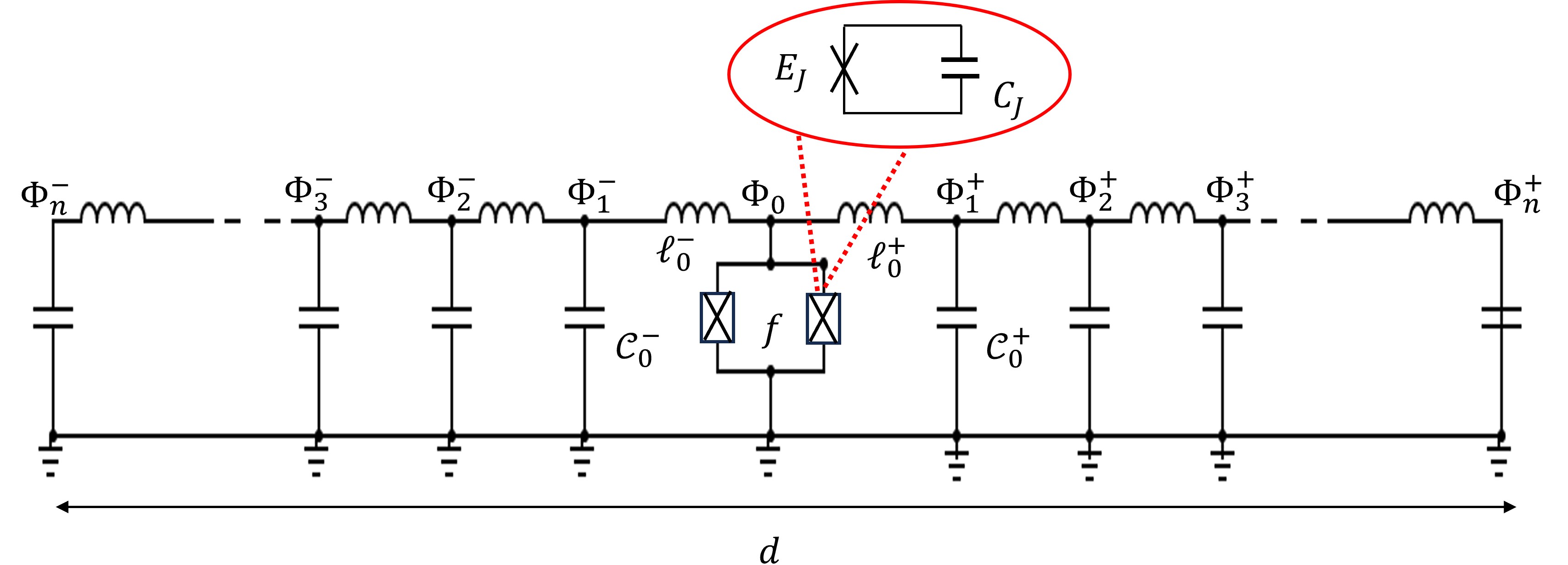}
    \caption{The schematics of the system consisting of a transmission line interrupted by a SQUID. Each transmission line element consists of a capacitance $\mathcal{C}_0^\pm$ and an inductance $\ell_0^\pm$. The SQUID itself consists of two Josephson junctions with Josephson energy $E_J$ in parallel to a capacitance $C_J$. The SQUID is threaded by a flux $f$. The quantities $\Phi_j^\pm$ denotes the fluxes at the nodes of the system. }\label{simple coupled circuits}
\end{figure*}
    
Within the framework of thermodynamics, it is of particular interest to explore desirable ways to transform energy. Quantum engines have been broadly studied \cite{uzdin2015equivalence,zhang2014quantum,rossnagel2014nanoscale,chand2017measurement,klaers2017squeezed,camati2019coherence,rossnagel2016single}, and it was found that in those settings work can either be extracted from thermal baths, or used to transfer heat from a cold source to a warmer one, thus working as a small-scale refrigerator.
However, quantum engines in general require measurements and external control over the parameters and interactions. This makes them more energetically costly than their classical counterparts when realizing them in practice. Therefore, autonomous quantum engines \cite{levy2012quantum,palao2001quantum,linden2010small}, which avoid the necessity of external control and are thus less demanding in terms of energy resources, are desirable.
In that respect, circuit QED has emerged as a promising platform for their study, since they allow a precise control over several parameters. For instance, Hofer \textit{et al.} propose to use a Josephson junction (JC) to create an interaction among three harmonic oscillators \cite{hofer_autonomous_2016} such that a three-body interaction Hamiltonian can be built, which creates an absorption refrigerator \cite{maslennikov_quantum_2019,nimmrichter_quantum_2017}.

Another important success of circuit QED was the first experimental realization of the Dynamical Casimir Effect (DCE) \cite{dodonov2020fifty}. The DCE is a phenomenon in which real photons are created out of the vacuum by the presence of time-dependent boundary conditions of the electromagnetic (EM) field \cite{lupascu_quantum_2005}. Given the difficulty of reaching the required frequencies using mechanical setups (e.g., in the conventional configuration consisting of one perfectly conducting cavity with one moving wall), the experimental circuit QED setup consisted of a cavity interrupted by a superconducting quantum interference device (SQUID) \cite{clarke_squid_2004}. SQUIDs are superconducting loops consisting of two Josephson junctions and are used to obtain precise measurements of magnetic fields. 
By providing an external driving field to the SQUID, it can mimic the boundary conditions that would be imposed by a fast oscillating mirror on the EM field, which creates pairs of photons inside the cavity \cite{wilson_observation_2011}. Building on this, similar setups have been proposed to exploit the effects of vacuum fluctuations, in particular to use the DCE as a resource to create entanglement between superconducting qubits \cite{felicetti_dynamical_2014}. The crucial element in these scenarios is the externally driven SQUID that creates an interaction between pairs of modes of the cavity. However, the degrees of freedom of the SQUID will be absent in the interaction and it only acts as a time-dependent phase in the Hamiltonian of the system.  

In the present work, we extend the previous studies by investigating the effect of including the internal degrees of freedom of the SQUID on the modes of the cavity with which it interacts.    
Our proposed setup, shown in Fig.~\ref{simple coupled circuits}, consists of two cavities connected by a SQUID in the middle. This can be interpreted as a transmission line interrupted by a SQUID \cite{wustmann_parametric_2013, fosco_vacuum_2013, lupascu_quantum_2005, devoret_circuit_2014, wallquist_selective_2006, bourassa_josephson_2012}. By quantizing the SQUID and the cavity field, we find that the SQUID degrees of freedom will generate a three-body interaction Hamiltonian with the cavity modes. We show that, by truncating the cavity to only two interacting modes, the DCE setup will reduce to an absorption refrigerator. Therefore, by assigning temperatures to the cavity and SQUID modes, one can use the SQUID as a work source to cool down the coldest mode of the cavity. 
Furthermore, since the SQUID couples to all the modes of the cavity, we investigate the effect of the interplay of more than one pair of modes on the performance of the cooling process. We observe that, depending on the temperature of the extra modes of the cavity, they can have a positive or a negative effect on the cooling. 

This paper is structured as follows. In Sec.~\ref{sec:setup} we present our setup by explaining the Lagrangian of the system. In Sec.~\ref{sec:lagrangian}, following Ref.~\cite{law_interaction_1995}, we derive the quantized Hamiltonian of the SQUID and the cavity. Employing this Hamiltonian, we show in Sec.~\ref{sec:refrigerator} that this setup can be reduced to an absorption refrigerator Hamiltonian. Finally, in Sec.~\ref{sec:enhanced_cooling}, we study the effect of adding more modes of the cavity on the cooling performance. We present our conclusions in Sec.~\ref{sec:Conclusions}.

\section{The system}\label{sec:setup}

The main object of study of this paper will be a system consisting of a transmission line with a SQUID in the middle, as illustrated in Fig.~\ref{simple coupled circuits}. This system is analogous to an optical cavity divided by a wall, which can in turn impose a time-dependent boundary condition on the EM field \cite{felicetti_dynamical_2014}.
 
 In our system, the cavities are represented by a large but finite number of LC circuits, each of them consisting of identical inductors and capacitors per unit length denoted by $\ell_0^{\pm}$ and $\mathcal{C}_0^{\pm}$, respectively. The primary variables used to describe the transmission lines are the node flux variables $\Phi^{\pm}_j(t)$ associated with each node in the cavity. The SQUID, which plays the role of the wall in the optical analog, consists of two parallel Josephson junctions (JJ) with identical Josephson energies $E_J$ and capacitances $C_J$. It is represented by the node flux at $x=0$, referred to as $\Phi_0$, and the phase difference $2f$, which is associated with the SQUID self-inductance $L$. Additionally, this phase difference can be affected by an external flux.
 
 We begin our analysis by first writing the Lagrangian of the cavities and the SQUID. The total Lagrangian of the system shown in Fig.~\ref{simple coupled circuits} can be written as 
 %
\begin{align}
    \L &= \L_c + \L_s + \L_f.
    \label{eq:discreet_Lagrang_cavity}
\end{align}
The Lagrangian $\L_c$ describes the bare cavity and is obtained by summing over the Lagrangians of each LC circuit element of the two cavities on the right ($\alpha = +$) and left ($\alpha = -$) side of the SQUID,
\begin{align}
    \L_c
&= 
    \frac{1}{2} \left(\frac{\hbar}{2e}\right)^2 \sum_{\alpha=\pm} \bigg[
    \sum_{j=1}^{N}  \Delta x \mathcal{C}^\alpha_0 (\dot{\Phi}_{j}^\alpha)^2  \notag \\
&-
    \sum_{j=0}^{N-1} \frac{\left(\Phi^\alpha_{j+1}-\Phi^\alpha_{j}\right)^2}{\ell_{0}^\alpha \Delta x}\bigg],
\end{align}
where $\dot{\Phi}_j^\alpha \equiv \partial_t \Phi^\alpha_j$, $\Phi_0^+ = \Phi_0^- \equiv \Phi_0$, and we have introduced $\Delta x$, the length of each circuit element. Next, the Lagrangian $\L_s$ pertains to the SQUID part of the system,
\begin{align}
    \L_s
&=
    \left(\frac{\hbar}{2e}\right)^2C_{J}\left(\dot{\Phi}^2_{0}+\dot{f}^{2}\right) 
+ 
    E_{J} \sum_{\alpha=\pm }\cos(\Phi_{0}+\alpha f).
\end{align}
We can identify the first term as the sum of the Coulomb energies in the SQUID. The second term denotes the Josephson energies which are obtained by assuming that the SQUID inductance $L$ is split into two equal parts $L/2$ shared between the two JJs, with a phase drop of $f$ over each part. Therefore, the phase difference on one of the JJs is $\Phi_{0}-f$ while it is $\Phi_{0}+f$ on the other one. 

Finally, the kinetic energy of the SQUID ring inductance $L$ corresponds to the term 
\begin{align}
    \L_f
&= - 
    \left(\frac{\hbar}{2e}\right)^2\frac{1}{2L}\left(4f^2+8\frac{M}{L_{\rm ext}}ff_{\rm ext}\right),
\end{align}
where the first term inside the bracket is the kinetic energy produced by the inductance $L$. In the second term we have accounted for the presence of an external circuit with inductance $L_{\rm ext}$ which is magnetically coupled to the SQUID by a mutual inductance $M$, which can be modulated by the external flux $f_{\rm ext}$.

Next, we perform a continuum limit by introducing the field,
\begin{align}
    \Phi(\alpha j \Delta x) \equiv \Phi^\alpha_{j}
\end{align}
for $j \in \{0, \ldots, N\}$ and $\alpha = \pm$, and by taking the limit $\Delta x \to 0$. 
The Lagrangian of the SQUID coupled to the cavity becomes in the continuum limit,
\begin{align}\label{eq:lagrangian_cav}
& \L_{\rm cav} = \L_c + \L_{s} = \notag \\
& \left(\frac{\hbar}{2e}\right)^2 \frac{\mathcal{C}_{0}}{2}\int_{-d/2}^{d/2}dx\left\{ \left[1+\frac{2C_J}{\mathcal{C}_0}\delta(x)\right]\dot{\Phi}^{2}-v^{2}\Phi'^{2} \right\}\notag\\
&+ E_J \int_{-d/2}^{d/2} dx \delta(x)\cos(\Phi)\cos(f).
\end{align}
where $\Phi' \equiv \partial_x \Phi$, $d = 2 N \Delta x$ denotes the lengths of the side cavities, and $v=1/\sqrt{\ell_0 \mathcal{C}_0}$ is the cavity field propagation velocity. Note that for simplicity we assumed identical cavities to the left and right, i.e., $\ell_0 = \ell_0^\alpha$ and $\mathcal{C}_0 = \mathcal{C}_0^\alpha$.

Next, we can employ the full Lagrangian to derive the equations of motion. Defining $E_L=\hbar^2/(8 e^2 L)$, we obtain the Euler-Lagrange equations for the variables $\Phi(x)$, $\Phi_0 = \Phi(0)$ and $f$,
\begin{gather}
\ddot{\Phi}-v^2\Phi{''}=0\label{eq_motion}\\
2\left(\frac{\hbar}{2e}\right)^2\!\!\left[C_J\ddot{\Phi}_{0}-\mathcal{C}_0 v^{2}(\Phi_{0+}^{\prime}-\Phi_{0-}^{\prime})\right]\!\!+\!2E_J\cos(f)\Phi_{0}=0\label{eq:eq_motion_1}\\
\left(\frac{\hbar}{2e}\right)^2C_J\ddot{f}+E_J\cos\Phi_0\sin f+E_L\left(f+\frac{M}{L_{ext}}f_{ext}\right)=0\label{eq:eq_motion_2}.
\end{gather}
These equations describe two coupled nonlinear oscillators, so the solution will in general feature chaotic behavior. With this in mind, we will restrict our analysis to the case in which the flux variable $\Phi_0\ll1$ and this is satisfied in the phase regime of superconducting qubits where $E_J\gg(2e)^2/(2C_J)$ \cite{devoret2004superconducting}. In this case, Eqs.~\eqref{eq:eq_motion_1} and \eqref{eq:eq_motion_2} will decouple, and the equation of motion for $f$ will read
\begin{equation}
   \left(\frac{\hbar}{2e}\right)^2 C_J\ddot{f}+E_J\sin f+E_L\left(f+\frac{M}{L_{ext}}f_{ext}\right)=0,
\end{equation}
and for small values of $f$ it will describe a shifted harmonic oscillator. 

Before further discussing the dynamics of $f$, we will focus on the equation of motion \eqref{eq_motion} for $\Phi(x,t)$. It can be solved by introducing a time-dependent basis \cite{fosco_vacuum_2013} such that
\begin{equation}\label{eq:separation}
    \Phi(x,t)=\sum_{n}\phi_n(t)\psi_{n}(x,t),
\end{equation}
where the summation runs over all the modes of the cavity and $\phi_n(t)$ is the time-dependent flux variable, while $\psi_n(x,t)$ represents the instantaneous eigenmodes of the cavity and obeys the equations
\begin{align}
    &  \psi_{n}''(x,t)+k^2_{n}(t)\psi_{n}(x,t) =0, \nonumber \\
    & \psi'_n(-d/2,t) = \psi'_n(d/2,t ) =0, \label{eq:boundary_deriv}\\
    & \psi_{n}'(0^{+},t)-\psi_{n}'(0^{-},t) = \nonumber \\
    & \frac{2}{v^{2}}\left(-\frac{C_{J}v^2k_n^2(t)}{\mathcal{C}_0}+\frac{E_{J}\cos f(t)}{\left(\frac{\hbar}{2e}\right)^2\mathcal{C}_0}\right)\psi_{n}(0,t), \nonumber 
\end{align}
where $k_n(t)=\omega_n(t)/v$ is the wave vector and ${\omega_n(t)\equiv \omega(f(t))}$ is the time-dependent frequency of the $n$th oscillator which can be obtained by solving Eq.~\eqref{eq:boundary_deriv} (see Ref.~\cite{velasco2022photon} for more details). It is important here to remark that the time-dependence in the functions $\psi_n(x,t)$ is a result of the time-dependence of the frequencies of the cavity modes $\omega_n(t)$, which is in turn caused by the time-dependent SQUID flux variable $f(t)$. Using Eq.~\eqref{eq:separation} and the given boundary conditions one can obtain the Lagrangian of the cavity $\L_{\rm cav}$ in terms of $\phi_n(t)$ and $\psi_n(x,t)$ in the following form (see App.~\ref{app:lagrangian} for more details),
\begin{align}
    \L_{\rm cav}
&=
    \frac{\hbar^2}{4E_C}\sum_{n}\left(\dot{\phi}_{n}^{2}-\omega_{n}^{2}(t)\phi_{n}^{2}\right)
    +\frac{\hbar^2 \dot{f}}{2E_C}\sum_{nm}M_{nm}\dot{\phi}_{n}\phi_{m}\notag\\
&+
    \frac{\hbar^2}{4E_C}\frac{\dot{f}^{2}}{2}\sum_{nmk}M_{nk}M_{mk}\phi_{n}\phi_{m}.\label{eq:lagrangian_fields}
\end{align}
In the above Lagrangian we defined $E_C=(2e)^2/(2C)$ as the charging energy corresponding to the capacitance $C$ of the cavity mode, and we have defined
\begin{equation}\label{eq:M}
    M_{nm}=\frac{1}{d}\int_{-d/2}^{d/2}dx [1+2C_{J}/\mathcal{C}_0\delta(x)] \psi_{m} \frac{d\psi_{n}}{df} \, ,
\end{equation}
where the time-dependence of the $\psi_n$ functions is related to the field $f$. The Lagrangian~\eqref{eq:lagrangian_fields} indicates that the time-dependence in $f$ causes an interaction among the modes of the cavity. We can interpret such an interaction as a manifestation of the Dynamical Casimir effect (DCE). In the standard DCE setting, the SQUID flux $f$ is externally driven, thus inducing an interaction among the cavity modes (two-mode squeezing) \cite{felicetti_dynamical_2014, wilson_observation_2011, drechsler2020state}.
In contrast, we are aiming at considering $f$ as an interacting mode rather than a time-dependent phase. To do so, we will consider the full Lagrangian of the system $\L=\L_{\rm cav}+\L_f$ to find the total Hamiltonian. Following a procedure analogous to Ref.~\cite{law_interaction_1995}, we will find the quantized Hamiltonian of the cavity and the SQUID in the next section.

%
%
%
\section{Quantized Hamiltonian }\label{sec:lagrangian}

We can write the full Lagrangian $\L$ of the system by combining Eq.~\eqref{eq:lagrangian_fields} with $\L_f$ from the SQUID Lagrangian,
\begin{align}
    \L
&=
    \frac{\hbar^2}{4E_C}\sum_{n}\left(\dot{\phi}_{n}^{2}-\omega_{n}^{2}(t)\phi_{n}^{2}\right)
    +\frac{\hbar^2}{2E_C}\dot{f}\sum_{nm}M_{nm}\dot{\phi_{n}}\phi_{m}\notag\\
&+
    \frac{\hbar^2}{4E_C}\frac{\dot{f}^{2}}{2}\sum_{n,m,k}M_{nk}M_{mk}\phi_{n}\phi_{m}
    +\frac{\hbar^2}{2E_{C_J}}\frac{\dot{f}^{2}}{2}-V(f),
\label{eq:lagrang_total}
\end{align}
where we have defined the energy of the Josephson capacitance as $E_{C_J}=(2e)^2/2C_J$, and
\begin{equation}
    V(f)=-E_{J}\cos f+\left(\frac{\hbar}{2e}\right)^2\frac{1}{L}\left(f^2+\frac{M}{L_{ext}}ff_{ext}\right)
\end{equation}
is the potential term.

From this Lagrangian one obtains the momentum variables conjugate to the SQUID flux $f$ and the cavity fluxes $\phi_n$,
\begin{align}
q_{n}&=\frac{1}{\hbar}\frac{\partial \L}{\partial\dot{\phi}_{n}}=\frac{\hbar}{2E_C}\left(\dot{\phi}_{n}+\dot{f}\sum_{m}M_{nm}\phi_{m}\right),\\
p_{f}&=\frac{1}{\hbar}\frac{\partial \L}{\partial\dot{f}}=\frac{\hbar}{2E_{C_J}}\dot{f}\\
&+\frac{\hbar}{2E_C}\left(\dot{f}\sum_{nmk}M_{nk}M_{mk}\phi_{n}\phi_{m}+\sum_{nm}M_{nm}\dot{\phi_{n}}\phi_{m}\right)\notag \, . 
\end{align}
Here, $q_n$ denotes the charge variable of the $n$th oscillator of the cavity and $p_f$ is the momentum variable of the SQUID. Note that the fraction $1/\hbar$ is added to make the momenta dimensionless. By performing a Legendre transformation of the Lagrangian we can find the Hamiltonian of the total system in terms of the flux and charge variables. Moreover, we promote the momentum and position variables to quantum operators by imposing the commutation relation $[\phi_n,q_n]=i$. In this way we can substitute
\begin{align}
       \phi_{n}&=\sqrt{\frac{\hbar\omega_{n}(t)}{2E_{C}}}\left(a_{n}+a_{n}^{\dagger}\right),\\
       q_{n}&=-i\sqrt{\frac{E_{C}}{2\hbar\omega_{n}(t)}}\left(a_{n}-a_{n}^{\dagger}\right).\label{eq:ladders_modes}
\end{align}
Hence, the Hamiltonian will take the following form,
\begin{gather}\label{eq:hamil_exact}
H=\sum_{n}\hbar\omega_{n}(t)a_{n}^{\dagger}a_{n}+\frac{\left(2e\right)^{2}}{2C_{J}}\left[p_{f}+\Gamma(f)\right]^{2}+V(f) \, ,
\end{gather}
where $\Gamma (f)\equiv (i/2)\sum_{n m}M_{nm}\left(a_{n}-a_{n}^{\dagger}\right)\left(a_{n}+a_{n}^{\dagger}\right)$. This Hamiltonian indicates that the SQUID momentum will interact with the cavity modes through the term $\Gamma$. The $f$ dependence of $\Gamma$ comes from the fact that the operators $a_n$ and $a^{\dagger}_n$ in \eqref{eq:ladders_modes} depend on $f$ through the time-dependent frequency $\omega_{n}(t)$.

At this point we are in a position to perform a first linear approximation of the Hamiltonian. This can be done by assuming that the position variable $f$ of the SQUID undergoes only small oscillations about its rest position $f_0$, i.e., $f(t)\approx f_{0}+\delta f(t)$ where $\delta f(t)\ll1$. In this case, the frequency can be expanded to first order in $\delta f$ as $\omega_{n}(f)\approx \omega_{n}(f_{0})+\delta f\omega_{n}^{\prime}(f_{0})$. Replacing this into Eq.~\eqref{eq:ladders_modes} we find, to first order in $\delta f$,
\begin{equation}\label{eq:anni_approx}
    a_{n}\approx a_{n}(f_{0})-\frac{1}{2}\delta f\frac{\omega_{n}^{\prime}(f_{0})}{\omega_{n}(f_{0})}a_{n}^{\dagger}(f_{0}) \, .
\end{equation}
Moreover, we should also do the same expansion for the other $f$-dependent parameters of the system, i.e., $M_{nm}$ and $\Gamma (f)$. Therefore, we write ${M_{nm}=M_{nm0}+\delta f M'_{nm}}$ with ${M'_{mn0}=\sum_{k}M_{nk0}M_{mk0}}$ and
\begin{equation}
   M_{nm0} =\frac{1}{d}\int_{-d/2}^{d/2}dx[1+2C_{J}/\mathcal{C}_0\delta(x)]\psi_{m}(f_{0})\frac{d\psi_{n}(f_0)}{df_0}.
\end{equation}
Using this we can also write 
$\Gamma(f)\approx\Gamma(f_0)+\delta f \Gamma'(f_0)$ such that 
\begin{gather}
    \Gamma'(f_0)=\sqrt{\frac{\omega_{m}(f_{0})}{\omega_{n}(f_{0})}}M_{nm}^{\prime}\left(a_{m}+a_{m}^{\dagger}\right)\left(a_{n}-a_{n}^{\dagger}\right)\notag\\
    +\sqrt{\frac{\omega_{m}(f_{0})}{\omega_{n}(f_{0})}}M_{nm}\frac{\omega_{n}^{\prime}(f_{0})}{\omega_n(f_0)}\left(a_{m}+a_{m}^{\dagger}\right)\left(a_{n}-a_{n}^{\dagger}\right),
\end{gather}
where we have dropped the $f_0$ dependence of the ladder operators. Now we can insert the above expansions into Eq.~\eqref{eq:hamil_exact} and find the Hamiltonian in terms of $\delta f$. Dropping the $f_0$ dependence of $\omega'_n$ and $\omega_n$ we find
\begin{gather}
    H=\hbar\sum_{n}\left(\omega_{n}+\omega_{n}^{\prime}\delta f\right)\left(\!\!a_{n}^{\dagger}\!-\delta f\frac{\omega_{n}^{\prime}}{2\omega_{n}}a_{n}\!\!\right)\left(\!\!a_{n}\!-\delta f\frac{\omega_{n}^{\prime}}{2\omega_{n}}a_{n}^{\dagger}\!\!\right)\notag\\
    +E_{C_J}\left(p_{f}+\Gamma_0+\delta f\Gamma'_0\right)^{2}+V(\delta f)
    \, .
    \label{eq:hamil_expand}
\end{gather}

Equation~\eqref{eq:hamil_expand} evidently denotes the nonlinear interaction between the modes and $\delta f$. To find the first linear order of the interaction Hamiltonian in terms of $\delta f$, we perform a unitary transformation $H'=T^{\dagger} H T$ where
\begin{equation}
    T=\exp\left\{ i\delta f(\Gamma_0+\frac{1}{2}\delta f\Gamma'_0)\right\}. 
\end{equation}
Applying this on the square term in Eq.~\eqref{eq:hamil_expand} we see that $ T^{\dagger} \left(p_{f}+\Gamma_0+\delta f\Gamma'_0\right)^{2} T=p_f^2$, so the transformation effectively shifts the momentum $p_f$ by $\Gamma_0+\delta\Gamma'_0$. The effect of the translation on the other parts of the Hamiltonian gives rise to interaction terms in powers of $\delta f$ (see App.~\ref{app:Hamiltonian} for details). To keep the linear interaction terms, we proceed to a rotating wave approximation (RWA) in order to eliminate fast oscillating terms in the Hamiltonian. To do so, we first promote $p_f$ and $\delta_f$ to quantum operators, as was done for the cavity variables, by defining their commutation relation $[\delta f,p_f]=i$. Thus, we can write
\begin{gather}
       \delta f=\sqrt{\frac{E_{C_{J}}}{\hbar\omega_{f}}}\left(a_{f}+a_{f}^{\dagger}\right) \, ,\\
       p_{f}=-i\frac{1}{2}\sqrt{\frac{\hbar\omega_{f}}{E_{C_{J}}}}\left(a_{f}-a_{f}^{\dagger}\right) \, ,
         \end{gather}
where 
\begin{align}
    \omega_f^2=\frac{E_{J}E_{C_{J}}}{\hbar^{2}}\left(\cos f_0+\left(\frac{\hbar}{2e}\right)^{2}\frac{2}{LE_{J}}\right)   
\end{align}
is the frequency of the oscillations in $\delta f$ (see App.~\ref{app:Hamiltonian} for details). To perform the RWA, we have to transform the Hamiltonian \eqref{eq:hamil_expand} into the rotating frame by the application of the unitary operator 
\begin{equation}
    U=\exp \left(\frac{it}{\hbar}\sum_{n}\hbar\omega_{n}a_{n}^{\dagger}a_{n}+\hbar\omega_{f}a_{f}^{\dagger}a_{f}\right).
\end{equation}
Afterwards we only keep the terms that satisfy the resonance condition $\omega_f=\omega_n+\omega_m$ for $m\neq n$. Going back to the Schr\"odinger picture, the Hamiltonian of the system becomes
\begin{align}
&    H_{\text{RWA}}\label{eq:hamil_final}
=
    \hbar\sum_{n}\omega_{n}a_{n}^{\dagger}a_{n}+\hbar\omega_{f}a_{f}^{\dagger}a_{f}\\
&-
    \frac{\hbar}{2}\sqrt{\frac{E_{C_{J}}}{\hbar\omega_{f}}}\sum_{n,m}M_{nm0}\sqrt{\frac{\omega_{m}}{\omega_{n}}}\!\omega_f\!\left(\!a_{f}^{\dagger}a_{m}a_{n}+a_{f}a_{m}^{\dagger}a_{n}^{\dagger}\!\right),\notag
\end{align}
where the summation runs over all the frequencies that satisfy the resonance condition. The Hamiltonian in Eq.~\eqref{eq:hamil_final} reflects the three-body interaction among the modes of the cavity and the SQUID mode. One interesting feature of this Hamiltonian is that by choosing only two interacting modes from the cavity, one can show that the system can behave as a quantum autonomous refrigerator.

\section{absorption refrigerator}\label{sec:refrigerator}
At this point, we proceed by simplifying the Hamiltonian~\eqref{eq:hamil_final} further by only considering two modes of the cavity, i.e., $n, m \in \{1,2\}$, so that the only interaction term contains $a_{f}^{\dagger}a_{2}a_{1} + \text{h.c.}$. The Hamiltonian then describes an \textit{absorption refrigerator} \cite{hofer_autonomous_2016,nimmrichter_quantum_2017,maslennikov_quantum_2019}. 

In this setup, the interaction term $a_{f}^{\dagger}a_{2}a_{1}$ entails that one photon is created in the SQUID while two photons are annihilated in the two modes of the cavity. This lowers the energy stored in the two modes and increases the energy of the SQUID, so that one can achieve autonomous cooling of one mode by utilizing the other two modes. 

On the other hand, the hermitian conjugate term $a_{f}a_{2}^{\dagger}a_{1}^{\dagger}$ describes a reverse process, i.e., one photon being annihilated in the SQUID while one photon is created at each cavity mode. However, one can favor the cooling term by conveniently choosing the energy levels of each oscillator. One possibility is to connect each cavity to a heat bath, allowing them to reach a thermal equilibrium state with temperatures $T_i$ for $i \in \{f, 1, 2\}$. By adjusting these temperatures, the occupation probabilities of the oscillators can be tuned. The initial occupation number $n_i$ of the $i$th oscillator can be expressed as
\begin{equation}
    n_i=\Tr[\rho_{T_i}a^{\dagger}_ia_i] \, ,
\end{equation}
where 
\begin{equation}
    \rho_{T_i}=\frac{\exp\left(-\frac{\hbar\omega_i}{k_B T_i}a^{\dagger}_ia_i\right)}{\Tr\left[\exp\left(-\frac{\hbar\omega_i}{ k_BT_i}a^{\dagger}_ia_i\right)\right]} 
\end{equation}
is the thermal initial state of the $i$th oscillator $i \in \{f, 1, 2\}$. 

In general, the system can operate both as a refrigerator and as a heat engine. By applying the resonance condition $\omega_f = \omega_1 + \omega_2$, one can adjust the temperatures of the three modes (corresponding to the initial occupation probabilities) in a way that allows the SQUID to cool down the other two cavities. This occurs when the temperature of the SQUID is higher than or equal to the cold mode ($T_2$) while still being smaller than the hot mode ($T_1$), i.e., $T_2 \leq T_f<T_1 $. In this temperature regime, the state of SQUID will have a lower occupation probability than to the other modes, so the final average energy of the SQUID, $E_f(t) = \mathrm{Tr}[\omega_f a_f^\dagger a_f \rho(t)]$, exceeds its initial energy obtained from the thermal state $E_f(t) = \mathrm{Tr}[\omega_f a_f^\dagger a_f \rho_{T_f}(0)]$. 
\begin{figure}
\centering
	\includegraphics[width=1.05\linewidth]{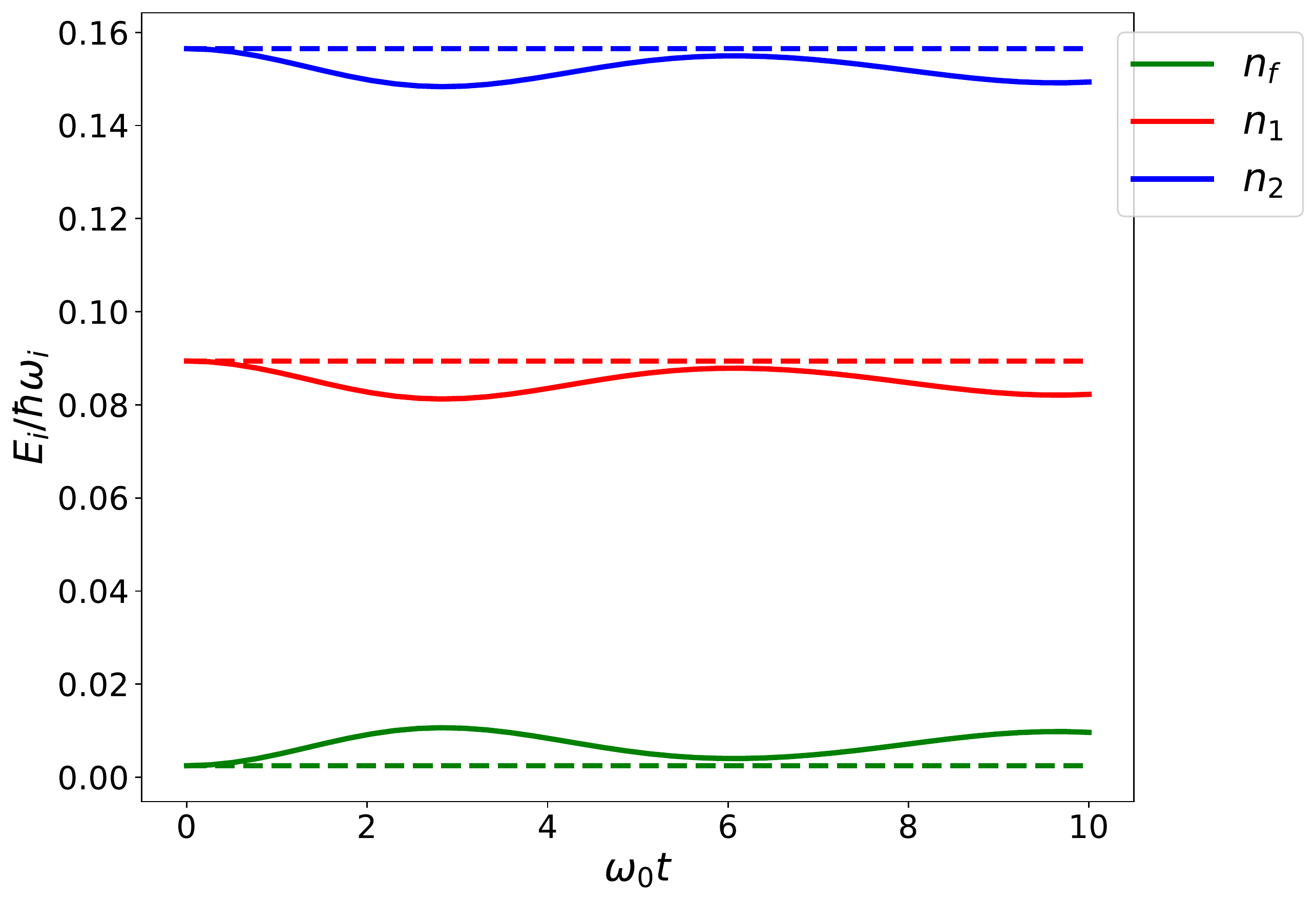}
	\caption{Averaged energies of each mode after time t, showing the increase in energy of the SQUID and the decrease in energy of the cold mode ($n_2$) and the hot mode ($n_1$). The solid lines show the average energy of each oscillator at time $t$. Dashed lines indicate the average energy at the initial time. This plot is depicted for frequencies are  $\omega_f=3\omega_0$, $\omega_2=2\omega_0$ and $\omega_1=\omega_0$ where $\omega_0=1$GHz and the temperatures are $T_f=65 (\hbar\omega_0/k_B)$, $T_2=T_f$ and $T_1=100 (\hbar\omega_0/k_B)$. }\label{fig:energetics}
\end{figure}
Moreover, the setup in Fig.~\ref{simple coupled circuits} offers the freedom to select the modes that are located on the left or right side of the SQUID. Therefore, it is possible to cool one half of the cavity by choosing parameters that localize the field modes \cite{velasco2022photon}. 
There are two options to ensure that the system will indeed work as a refrigerator. The first approach involves connecting each mode to separate heat baths initially, allowing them to reach their respective temperatures. Subsequently, the modes interact with each other while still being influenced by the baths. In this scenario, the dynamics at time $t$ can be obtained by solving a master equation as demonstrated in Ref.~\cite{hofer_autonomous_2016}.
An alternative approach is to disconnect the modes from the heat baths once they have reached their respective temperatures. Afterwards, the modes interact with each other without the influence of the baths \cite{nimmrichter_quantum_2017}. In this case, the dynamics are determined by the unitary time evolution. 
Since both cases lead to similar results, for simplicity and the sole purpose of demonstrating the behavior of the cooling system in our setup, we will consider the latter scenario from this point on. 

To obtain numerical results, we use the QuTip library \cite{johansson2012qutip} to solve the unitary dynamics caused by the rotated Hamiltonian~\eqref{eq:hamil_final}. In Fig.~\ref{fig:energetics}, we show the resulting average energy of each mode as a function of time $t$. The comparison with their initial values clearly indicates that the energy of the SQUID increases while the other two modes become colder. This confirms that this configuration operates as a refrigerator, consistent with the findings in Ref.~\cite{nimmrichter_quantum_2017,maslennikov_quantum_2019}. As a matter of fact, a better indication that this setup can work as a cooler can be achieved by monitoring the temperature of the cold mode while it remains in contact with a heat bath as shown in Ref.~\cite{hofer_autonomous_2016}. This involves solving a quantum master equation to find the dynamics of each mode. In contrast, here we limited ourselves to a closed system case as it still captures the essential dynamical behaviour of the cooler similarly to the open dynamics \cite{nimmrichter_quantum_2017}.

\section{Enhanced cooling }\label{sec:enhanced_cooling}

In the previous section, we show that by keeping only two interacting modes of the cavity, the system can act as an absorption refrigerator. However, one of the advantages of the proposed setup is that it gives us the freedom to retain more modes as long as they fulfill the resonance condition $\omega_f=\omega_n+\omega_m$. In this section we will study the effect on the refrigeration process of two extra modes from the cavity. 
\begin{figure}[t]
\centering
	\includegraphics[width=1.05\linewidth]{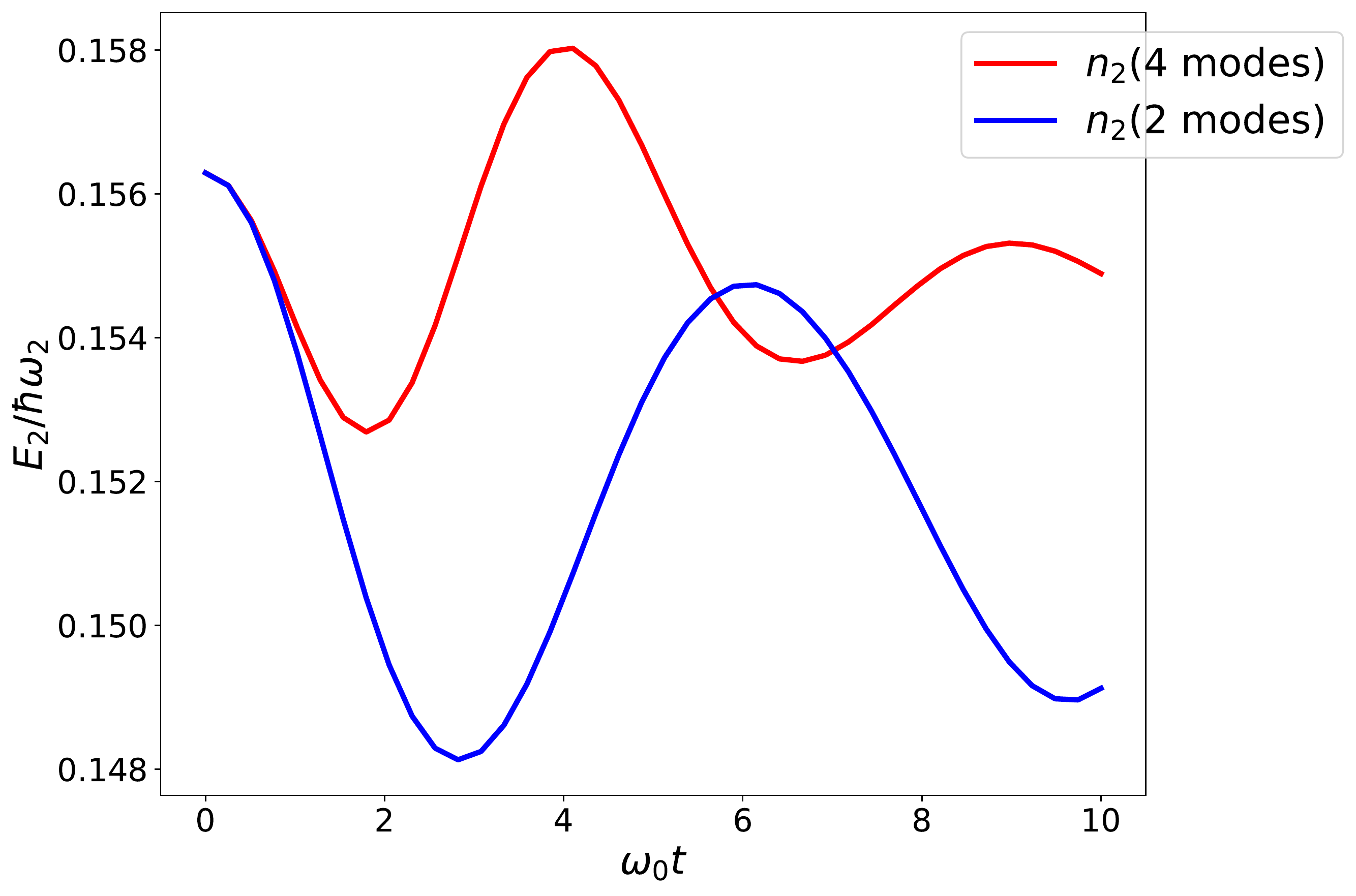}
	\caption{Average energy of the cold mode ($n_2$) as a function of time $t$. The blue line indicates the energy of the cold mode when the SQUID is interacting only with the hot and cold modes (two modes) of the cavity. The red line denotes the energy of the cold mode when two extra modes are added. The frequencies are $\omega_f=3\omega_0$, $\omega_3=1.8\omega_0$ and $\omega_4=1.2\omega_0$ and the temperatures are $T_f=65(\hbar\omega_0/k_B)$, $T_3=T_4=92(\hbar\omega_0/k_B)$. }\label{fig:comparison_high_temp}
\end{figure}

We now consider the Hamiltonian~\eqref{eq:hamil_final} with four modes $\omega_{1,2,3,4}$, such that they satisfy $\omega_f=\omega_3+\omega_4=\omega_1+\omega_2$. The resulting Hamiltonian will take the form 
\begin{align}\label{eq:hamil_four_modes}
&   H_{\rm ref}=\hbar\sum_{n=1,2}\omega_{n}a_{n}^{\dagger}a_{n}+\hbar\omega_{f}a_{f}^{\dagger}a_{f}\\
&-\frac{1}{2}\sqrt{\hbar\omega_fE_{C_{J}}}\sum_{n,m=1,2}\!\!M_{nm0}\sqrt{\frac{\omega_{m}}{\omega_{n}}}\left(a_{f}^{\dagger}a_{2}a_{1}+a_{f}a_{2}^{\dagger}a_{1}^{\dagger}\right)\notag\\
&-\frac{1}{2}\sqrt{\hbar\omega_fE_{C_{J}}}\sum_{n,m=3,4}\!\!M_{nm0}\sqrt{\frac{\omega_{m}}{\omega_{n}}}\left(a_{f}^{\dagger}a_{3}a_{4}+a_{f}a_{3}^{\dagger}a_{4}^{\dagger}\right)\notag.
\end{align}
\begin{figure}[t]
\centering
	\includegraphics[width=1.05\linewidth]{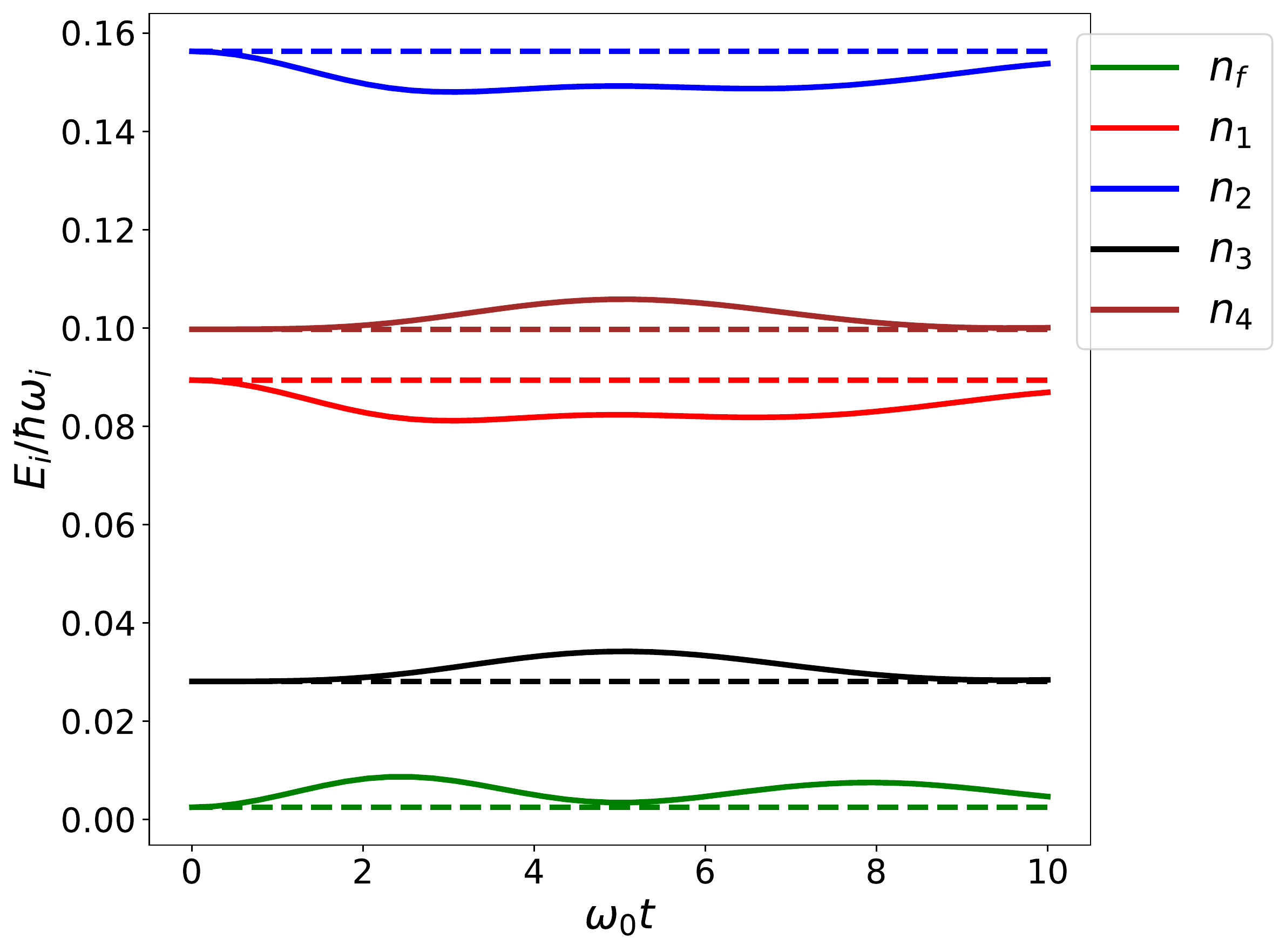}
	\caption{Averaged energy of all the modes after time $\omega_0 t$. The cold mode ($n_2$) cools down and the cooling period is longer compared to the case of only two interacting modes (see Fig.~\ref{fig:energetics}).
 }\label{fig:four_modes}
\end{figure}
The modes 1 and 2 are the hot and cold modes of the refrigerators, and they will not interact directly with the extra modes 3 and 4. To study the impact of these modes on the refrigeration process, we assign thermal states to the two modes with temperatures $T_3$ and $T_4$. 

It turns out that the refrigeration process will depend strongly on the temperatures of those modes. If we set their temperatures to be higher than the temperature of the cold mode ($T_2$) and the SQUID mode ($T_f$), the energy of the cold mode will be higher compared to the two-mode case. The reason lies behind the fact that when the SQUID mode interacts with the extra modes, due to their higher temperature, the SQUID mode will become hotter and its state population will grow. Therefore, since the SQUID is simultaneously in contact with the cold mode, due to its higher energy, it will not be able to decrease the energy of the cold mode as it did before. The result of this process is sketched in Fig.~\ref{fig:comparison_high_temp}. We notice, as expected, that when the temperatures of the two extra modes are higher than that of the SQUID and the cold mode, the cooling process is less effective. 

\begin{figure}[t]
\centering
	\includegraphics[width=1.05\linewidth]{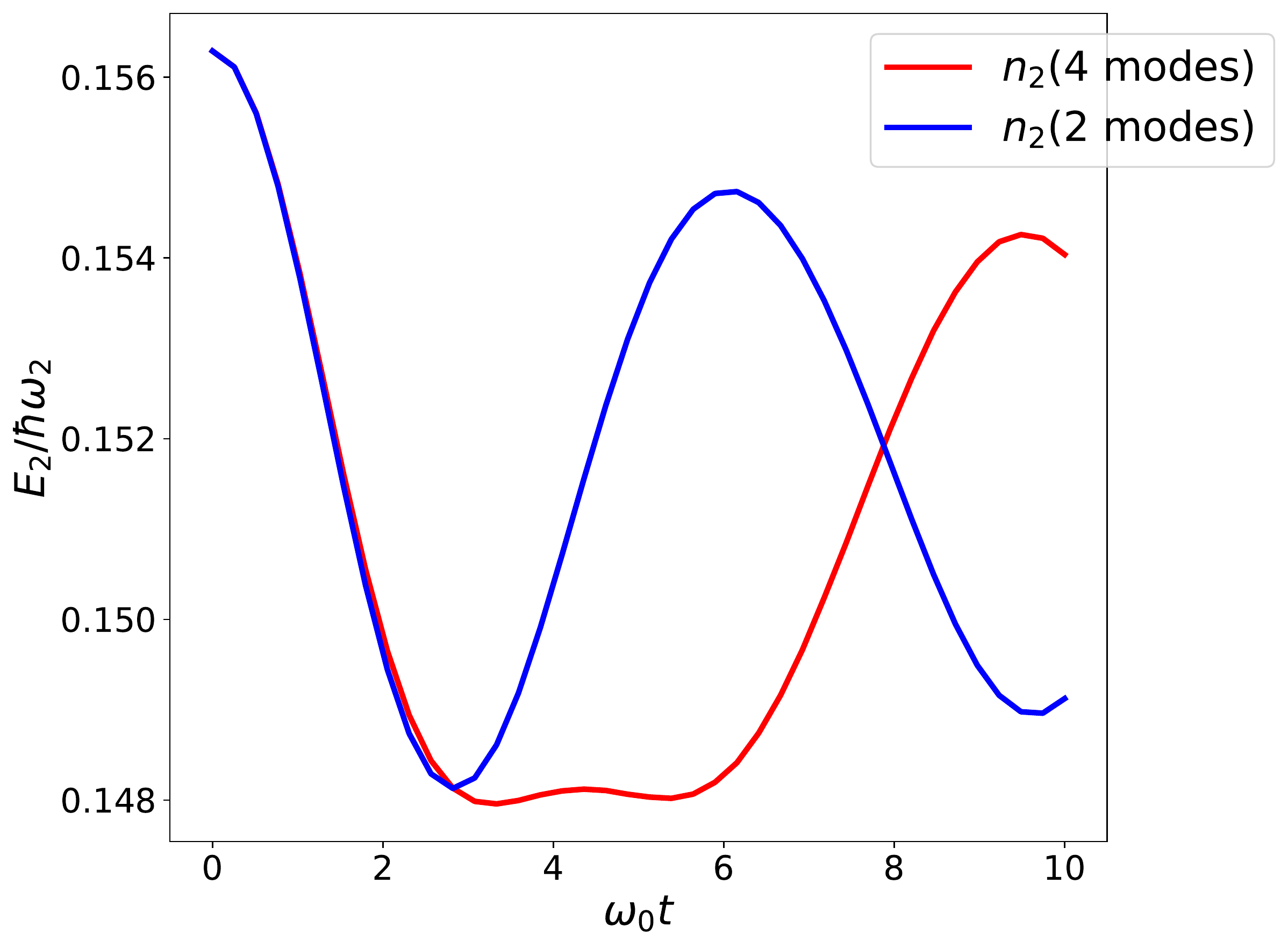}
	\caption{Averaged energy of the cold mode. When the SQUID is interacting with the two extra modes which have the same temperature of the cold mode, the cooling process is enhanced (red line). The frequencies are $\omega_f=3\omega_0$, $\omega_3=1.8\omega_0$ and $\omega_4=1.2\omega_0$, and the temperatures are $T_f=65 (\hbar\omega_f/k_B)$ and $T_3=T_4=T_f$. }\label{fig:comparison_low_temp}
\end{figure}
\begin{figure}[t]
\centering
	\includegraphics[width=1.06\linewidth]{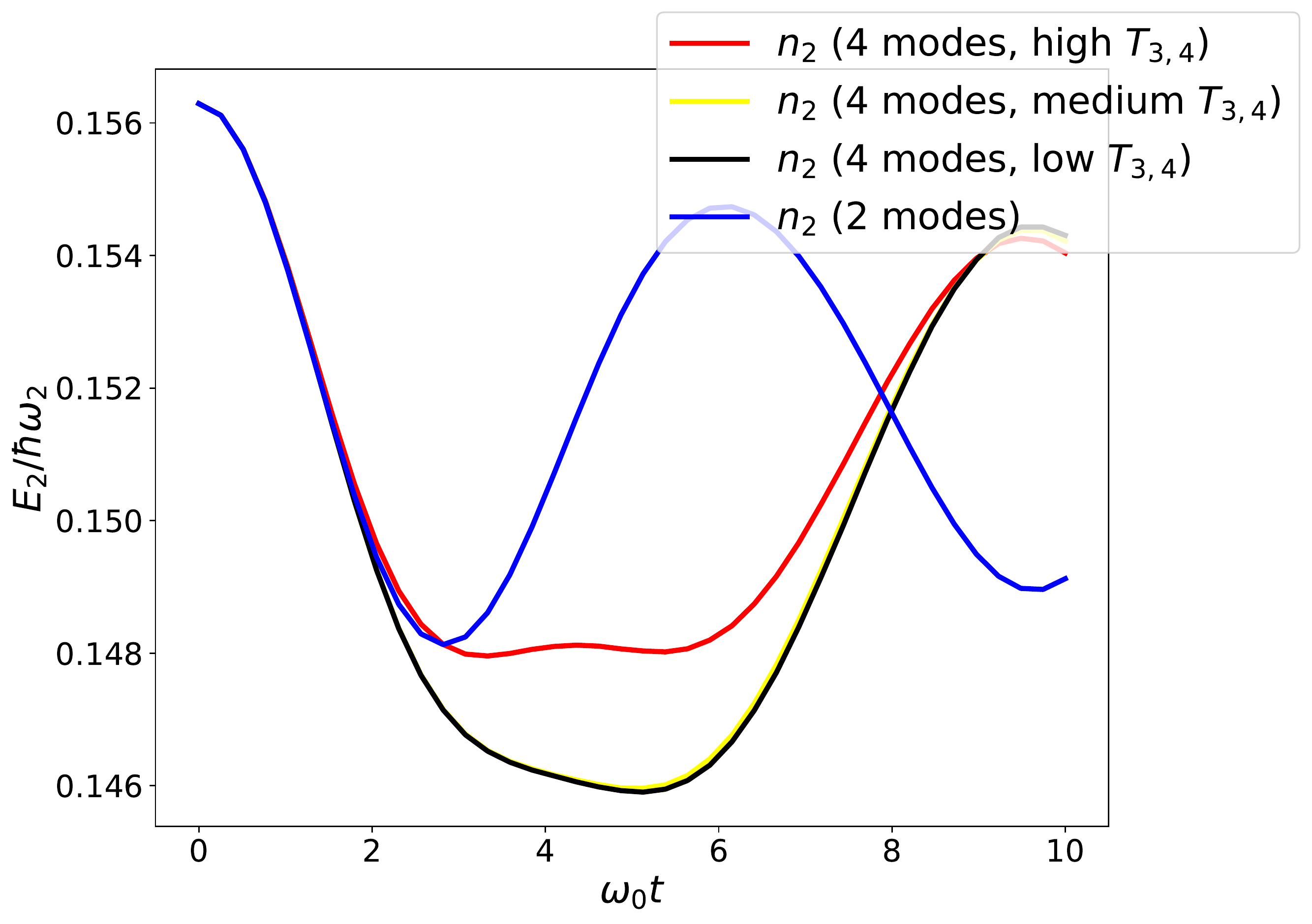}
	\caption{Averaged energy of the cold mode. By decreasing the temperature of the extra modes, one can further enhance the cooling of the cold mode ($n_2$). We compare the cooling in the case of two modes (blue line) to the case of cooling in the presence of four modes for $T_3 = T_4 = 65 \hbar \omega_0$ (red line), $T_3 = T_4 = 40 \hbar \omega_0$ (yellow line), and $T_3 = T_4 = \hbar \omega_0$ (black line). }\label{fig:temps}
\end{figure}
On the other hand, if we choose the temperatures of the extra modes to be equal to the temperatures of the SQUID and the cold mode, we observe enhanced cooling (see Fig.~\ref{fig:four_modes}). When the SQUID interacts with the cold mode, its energy will increase. Therefore, during the interaction with the extra modes, it will release its energy to those modes. Thus, when it comes back to interact with the cold mode again, it has a lower population than before and therefore it can lower the cold mode energy. This results in an enhancement in the refrigeration process, and the two extra modes are acting like heat sinks that decrease the energy of the SQUID during the interaction. Figure~\ref{fig:comparison_low_temp} shows the behavior of the cold mode when the temperatures of the extra modes are equal to that of the SQUID and the cold mode. We notice that this not only improves the cooling process but also that the cold mode keeps its minimum energy for a longer period of time before moving into the oscillatory dynamics.

The effect of decreasing the temperatures of the extra modes below the temperature of the SQUID and cold mode is also depicted in Fig.~\ref{fig:temps}. We observe that if we decrease the temperature of the extra modes to below the temperature of the cold and SQUID modes, we can improve the cooling effect even further. Nevertheless, this cooling will stop at some point which can be related to the fundamental limit of cooling \cite{freitas2017fundamental}. This effect can be observed in externally driven systems and we do not have drive in our Hamiltonian due to quantizing the SQUID degrees of freedom. However, since the original setup stems from the dynamical Casimir effect, we can argue that this cooling limit should still exist because we are considering the quantized version of the driven SQUID. This means that a time-dependent classical phase in the Hamiltonian is now treated quantum mechanically to create a three body Hamiltonian instead of two body interaction.

\section{Conclusions}\label{sec:Conclusions}

We investigated cooling in a setup consisting of a finite transmission line interrupted by a SQUID, a system that has been used to experimentally realize the Dynamical Casimir effect \cite{wilson_observation_2011}. 
By quantizing the SQUID degrees of freedom, we obtained the Lagrangian of the quantized system in terms of the modes inside the cavity. After switching to the Hamiltonian representation, we obtained a three-body interaction term consisting of the cavity modes operators and the SQUID. By keeping only two interacting modes of the cavity we reduced the Hamiltonian into the form of the absorption refrigerator interaction. Therefore, we showed that in some temperature regime, by increasing the energy of the SQUID mode, one can cool down the two modes of the cavity.

To study the effect of the other existing modes of the cavity on the cooling process, we explored the influence of including two extra modes in the interaction with the SQUID. In this way we observed that, if the initial temperature of the extra modes are equal to the cold mode and the SQUID, one can cool the cold mode even further than the previous case. This result is significant as it demonstrates the versatility of the architecture depicted in Fig.~\ref{simple coupled circuits} where the system offers the flexibility to work with multiple cavity modes and their interactions with the SQUID. 

Since the Hamiltonian~\eqref{eq:hamil_final} contains a squeezing interaction term which stems from the DCE, we expect that this interaction among the cavity modes and the SQUID can create entanglement among the involved modes which may play a role in the cooling effect. Analyzing the role of entanglement for cooling is beyond the scope of the present work but would be interesting to study in future work.

\section{Acknowledgements}

The authors would like to thank J. N. Freitas for helpful discussions, and acknowledge financial support from the National
Research Fund Luxembourg under Grants C20/MS/14757511/OpenTop and C18/MS/12704391/QUTHERM, the Agencia Nacional de Promoción Científica y Tecnológica (ANPCyT), Consejo Nacional de Investigaciones Científicas y Técnicas (CONICET), and Universidad de Buenos Aires (UBA).
 
\bibliographystyle{apsrev4-2}
\bibliography{draft}

\begin{thebibliography}{35}%
\makeatletter
\providecommand \@ifxundefined [1]{%
 \@ifx{#1\undefined}
}%
\providecommand \@ifnum [1]{%
 \ifnum #1\expandafter \@firstoftwo
 \else \expandafter \@secondoftwo
 \fi
}%
\providecommand \@ifx [1]{%
 \ifx #1\expandafter \@firstoftwo
 \else \expandafter \@secondoftwo
 \fi
}%
\providecommand \natexlab [1]{#1}%
\providecommand \enquote  [1]{``#1''}%
\providecommand \bibnamefont  [1]{#1}%
\providecommand \bibfnamefont [1]{#1}%
\providecommand \citenamefont [1]{#1}%
\providecommand \href@noop [0]{\@secondoftwo}%
\providecommand \href [0]{\begingroup \@sanitize@url \@href}%
\providecommand \@href[1]{\@@startlink{#1}\@@href}%
\providecommand \@@href[1]{\endgroup#1\@@endlink}%
\providecommand \@sanitize@url [0]{\catcode `\\12\catcode `\$12\catcode
  `\&12\catcode `\#12\catcode `\^12\catcode `\_12\catcode `\%12\relax}%
\providecommand \@@startlink[1]{}%
\providecommand \@@endlink[0]{}%
\providecommand \url  [0]{\begingroup\@sanitize@url \@url }%
\providecommand \@url [1]{\endgroup\@href {#1}{\urlprefix }}%
\providecommand \urlprefix  [0]{URL }%
\providecommand \Eprint [0]{\href }%
\providecommand \doibase [0]{https://doi.org/}%
\providecommand \selectlanguage [0]{\@gobble}%
\providecommand \bibinfo  [0]{\@secondoftwo}%
\providecommand \bibfield  [0]{\@secondoftwo}%
\providecommand \translation [1]{[#1]}%
\providecommand \BibitemOpen [0]{}%
\providecommand \bibitemStop [0]{}%
\providecommand \bibitemNoStop [0]{.\EOS\space}%
\providecommand \EOS [0]{\spacefactor3000\relax}%
\providecommand \BibitemShut  [1]{\csname bibitem#1\endcsname}%
\let\auto@bib@innerbib\@empty
\bibitem [{\citenamefont {Devoret}\ \emph {et~al.}(1995)\citenamefont {Devoret}
  \emph {et~al.}}]{devoret1995quantum}%
  \BibitemOpen
  \bibfield  {author} {\bibinfo {author} {\bibfnamefont {M.~H.}\ \bibnamefont
  {Devoret}} \emph {et~al.},\ }\href
  {https://doi.org/https://boulderschool.yale.edu/sites/default/files/files/devoret_quantum_fluct_les_houches.pdf}
  {\bibfield  {journal} {\bibinfo  {journal} {Les Houches, Session LXIII}\
  }\textbf {\bibinfo {volume} {7}},\ \bibinfo {pages} {133} (\bibinfo {year}
  {1995})}\BibitemShut {NoStop}%
\bibitem [{\citenamefont {Girvin}()}]{devoret_circuit_2014}%
  \BibitemOpen
  \bibfield  {author} {\bibinfo {author} {\bibfnamefont {S.~M.}\ \bibnamefont
  {Girvin}},\ }in\ \href
  {https://doi.org/10.1093/acprof:oso/9780199681181.003.0003} {\emph {\bibinfo
  {booktitle} {Quantum Machines: Measurement and Control of Engineered Quantum
  Systems}}},\ \bibinfo {editor} {edited by\ \bibinfo {editor} {\bibfnamefont
  {M.}~\bibnamefont {Devoret}}, \bibinfo {editor} {\bibfnamefont
  {B.}~\bibnamefont {Huard}}, \bibinfo {editor} {\bibfnamefont
  {R.}~\bibnamefont {Schoelkopf}},\ and\ \bibinfo {editor} {\bibfnamefont
  {L.~F.}\ \bibnamefont {Cugliandolo}}}\ (\bibinfo  {publisher} {Oxford
  University {PressOxford}})\ \bibinfo {edition} {1st}\ ed.,\ pp.\ \bibinfo
  {pages} {113--256}\BibitemShut {NoStop}%
\bibitem [{\citenamefont {Devoret}\ \emph {et~al.}(2004)\citenamefont
  {Devoret}, \citenamefont {Wallraff},\ and\ \citenamefont
  {Martinis}}]{devoret2004superconducting}%
  \BibitemOpen
  \bibfield  {author} {\bibinfo {author} {\bibfnamefont {M.~H.}\ \bibnamefont
  {Devoret}}, \bibinfo {author} {\bibfnamefont {A.}~\bibnamefont {Wallraff}},\
  and\ \bibinfo {author} {\bibfnamefont {J.~M.}\ \bibnamefont {Martinis}},\
  }\bibfield  {journal} {\bibinfo  {journal} {arXiv preprint cond-mat/0411174}\
  }\href {https://doi.org/https://arxiv.org/abs/cond-mat/0411174}
  {https://arxiv.org/abs/cond-mat/0411174} (\bibinfo {year} {2004})\BibitemShut
  {NoStop}%
\bibitem [{\citenamefont {Beaudoin}\ \emph {et~al.}(2011)\citenamefont
  {Beaudoin}, \citenamefont {Gambetta},\ and\ \citenamefont
  {Blais}}]{beaudoin2011dissipation}%
  \BibitemOpen
  \bibfield  {author} {\bibinfo {author} {\bibfnamefont {F.}~\bibnamefont
  {Beaudoin}}, \bibinfo {author} {\bibfnamefont {J.~M.}\ \bibnamefont
  {Gambetta}},\ and\ \bibinfo {author} {\bibfnamefont {A.}~\bibnamefont
  {Blais}},\ }\href {https://doi.org/10.1088/0031-8949/2009/T137/014012}
  {\bibfield  {journal} {\bibinfo  {journal} {Physical Review A}\ }\textbf
  {\bibinfo {volume} {84}},\ \bibinfo {pages} {043832} (\bibinfo {year}
  {2011})}\BibitemShut {NoStop}%
\bibitem [{\citenamefont {Majer}\ \emph {et~al.}(2007)\citenamefont {Majer},
  \citenamefont {Chow}, \citenamefont {Gambetta}, \citenamefont {Koch},
  \citenamefont {Johnson}, \citenamefont {Schreier}, \citenamefont {Frunzio},
  \citenamefont {Schuster}, \citenamefont {Houck}, \citenamefont {Wallraff}
  \emph {et~al.}}]{majer2007coupling}%
  \BibitemOpen
  \bibfield  {author} {\bibinfo {author} {\bibfnamefont {J.}~\bibnamefont
  {Majer}}, \bibinfo {author} {\bibfnamefont {J.}~\bibnamefont {Chow}},
  \bibinfo {author} {\bibfnamefont {J.}~\bibnamefont {Gambetta}}, \bibinfo
  {author} {\bibfnamefont {J.}~\bibnamefont {Koch}}, \bibinfo {author}
  {\bibfnamefont {B.}~\bibnamefont {Johnson}}, \bibinfo {author} {\bibfnamefont
  {J.}~\bibnamefont {Schreier}}, \bibinfo {author} {\bibfnamefont
  {L.}~\bibnamefont {Frunzio}}, \bibinfo {author} {\bibfnamefont
  {D.}~\bibnamefont {Schuster}}, \bibinfo {author} {\bibfnamefont {A.~A.}\
  \bibnamefont {Houck}}, \bibinfo {author} {\bibfnamefont {A.}~\bibnamefont
  {Wallraff}}, \emph {et~al.},\ }\href
  {https://doi.org/https://www.nature.com/articles/nature06184} {\bibfield
  {journal} {\bibinfo  {journal} {Nature}\ }\textbf {\bibinfo {volume} {449}},\
  \bibinfo {pages} {443} (\bibinfo {year} {2007})}\BibitemShut {NoStop}%
\bibitem [{\citenamefont {Wustmann}\ and\ \citenamefont
  {Shumeiko}()}]{wustmann_parametric_2013}%
  \BibitemOpen
  \bibfield  {author} {\bibinfo {author} {\bibfnamefont {W.}~\bibnamefont
  {Wustmann}}\ and\ \bibinfo {author} {\bibfnamefont {V.}~\bibnamefont
  {Shumeiko}},\ }\href {https://doi.org/10.1103/PhysRevB.87.184501} {\bibinfo
  {title} {Parametric resonance in tunable superconducting cavities}},\ \Eprint
  {https://arxiv.org/abs/1302.3484 [cond-mat, physics:quant-ph]} {1302.3484
  [cond-mat, physics:quant-ph]} \BibitemShut {NoStop}%
\bibitem [{\citenamefont {Fosco}\ \emph {et~al.}()\citenamefont {Fosco},
  \citenamefont {Lombardo},\ and\ \citenamefont
  {Mazzitelli}}]{fosco_vacuum_2013}%
  \BibitemOpen
  \bibfield  {author} {\bibinfo {author} {\bibfnamefont {C.~D.}\ \bibnamefont
  {Fosco}}, \bibinfo {author} {\bibfnamefont {F.~C.}\ \bibnamefont
  {Lombardo}},\ and\ \bibinfo {author} {\bibfnamefont {F.~D.}\ \bibnamefont
  {Mazzitelli}},\ }\href {https://doi.org/10.1103/PhysRevD.87.105008}
  {\bibfield  {journal} {\bibinfo  {journal} {Physical Review D}\ }\textbf
  {\bibinfo {volume} {87}},\ \bibinfo {pages} {105008}}\BibitemShut {NoStop}%
\bibitem [{\citenamefont {Senior}\ \emph {et~al.}(2020)\citenamefont {Senior},
  \citenamefont {Gubaydullin}, \citenamefont {Karimi}, \citenamefont
  {Peltonen}, \citenamefont {Ankerhold},\ and\ \citenamefont
  {Pekola}}]{senior2020heat}%
  \BibitemOpen
  \bibfield  {author} {\bibinfo {author} {\bibfnamefont {J.}~\bibnamefont
  {Senior}}, \bibinfo {author} {\bibfnamefont {A.}~\bibnamefont {Gubaydullin}},
  \bibinfo {author} {\bibfnamefont {B.}~\bibnamefont {Karimi}}, \bibinfo
  {author} {\bibfnamefont {J.~T.}\ \bibnamefont {Peltonen}}, \bibinfo {author}
  {\bibfnamefont {J.}~\bibnamefont {Ankerhold}},\ and\ \bibinfo {author}
  {\bibfnamefont {J.~P.}\ \bibnamefont {Pekola}},\ }\href
  {https://doi.org/https://www.nature.com/articles/s42005-020-0307-5}
  {\bibfield  {journal} {\bibinfo  {journal} {Communications Physics}\ }\textbf
  {\bibinfo {volume} {3}},\ \bibinfo {pages} {40} (\bibinfo {year}
  {2020})}\BibitemShut {NoStop}%
\bibitem [{\citenamefont {Karimi}\ and\ \citenamefont
  {Pekola}(2016)}]{karimi2016otto}%
  \BibitemOpen
  \bibfield  {author} {\bibinfo {author} {\bibfnamefont {B.}~\bibnamefont
  {Karimi}}\ and\ \bibinfo {author} {\bibfnamefont {J.}~\bibnamefont
  {Pekola}},\ }\href
  {https://doi.org/https://pubs.aip.org/aip/apl/article-abstract/68/14/1996/65614}
  {\bibfield  {journal} {\bibinfo  {journal} {Physical Review B}\ }\textbf
  {\bibinfo {volume} {94}},\ \bibinfo {pages} {184503} (\bibinfo {year}
  {2016})}\BibitemShut {NoStop}%
\bibitem [{\citenamefont {Kadijani}\ \emph {et~al.}(2020)\citenamefont
  {Kadijani}, \citenamefont {Schmidt}, \citenamefont {Esposito},\ and\
  \citenamefont {Freitas}}]{kadijani2020heat}%
  \BibitemOpen
  \bibfield  {author} {\bibinfo {author} {\bibfnamefont {S.~S.}\ \bibnamefont
  {Kadijani}}, \bibinfo {author} {\bibfnamefont {T.~L.}\ \bibnamefont
  {Schmidt}}, \bibinfo {author} {\bibfnamefont {M.}~\bibnamefont {Esposito}},\
  and\ \bibinfo {author} {\bibfnamefont {N.}~\bibnamefont {Freitas}},\ }\href
  {https://doi.org/https://doi.org/10.1103/PhysRevB.102.235422} {\bibfield
  {journal} {\bibinfo  {journal} {Physical Review B}\ }\textbf {\bibinfo
  {volume} {102}},\ \bibinfo {pages} {235422} (\bibinfo {year}
  {2020})}\BibitemShut {NoStop}%
\bibitem [{\citenamefont {Uzdin}\ \emph {et~al.}(2015)\citenamefont {Uzdin},
  \citenamefont {Levy},\ and\ \citenamefont {Kosloff}}]{uzdin2015equivalence}%
  \BibitemOpen
  \bibfield  {author} {\bibinfo {author} {\bibfnamefont {R.}~\bibnamefont
  {Uzdin}}, \bibinfo {author} {\bibfnamefont {A.}~\bibnamefont {Levy}},\ and\
  \bibinfo {author} {\bibfnamefont {R.}~\bibnamefont {Kosloff}},\ }\href
  {https://doi.org/10.1103/PhysRevX.5.031044} {\bibfield  {journal} {\bibinfo
  {journal} {Physical Review X}\ }\textbf {\bibinfo {volume} {5}},\ \bibinfo
  {pages} {031044} (\bibinfo {year} {2015})}\BibitemShut {NoStop}%
\bibitem [{\citenamefont {Zhang}\ \emph {et~al.}(2014)\citenamefont {Zhang},
  \citenamefont {Bariani},\ and\ \citenamefont {Meystre}}]{zhang2014quantum}%
  \BibitemOpen
  \bibfield  {author} {\bibinfo {author} {\bibfnamefont {K.}~\bibnamefont
  {Zhang}}, \bibinfo {author} {\bibfnamefont {F.}~\bibnamefont {Bariani}},\
  and\ \bibinfo {author} {\bibfnamefont {P.}~\bibnamefont {Meystre}},\ }\href
  {https://doi.org/10.1103/PhysRevLett.112.150602} {\bibfield  {journal}
  {\bibinfo  {journal} {Physical review letters}\ }\textbf {\bibinfo {volume}
  {112}},\ \bibinfo {pages} {150602} (\bibinfo {year} {2014})}\BibitemShut
  {NoStop}%
\bibitem [{\citenamefont {Ro{\ss}nagel}\ \emph {et~al.}(2014)\citenamefont
  {Ro{\ss}nagel}, \citenamefont {Abah}, \citenamefont {Schmidt-Kaler},
  \citenamefont {Singer},\ and\ \citenamefont {Lutz}}]{rossnagel2014nanoscale}%
  \BibitemOpen
  \bibfield  {author} {\bibinfo {author} {\bibfnamefont {J.}~\bibnamefont
  {Ro{\ss}nagel}}, \bibinfo {author} {\bibfnamefont {O.}~\bibnamefont {Abah}},
  \bibinfo {author} {\bibfnamefont {F.}~\bibnamefont {Schmidt-Kaler}}, \bibinfo
  {author} {\bibfnamefont {K.}~\bibnamefont {Singer}},\ and\ \bibinfo {author}
  {\bibfnamefont {E.}~\bibnamefont {Lutz}},\ }\href
  {https://doi.org/10.1103/PhysRevLett.112.030602} {\bibfield  {journal}
  {\bibinfo  {journal} {Physical review letters}\ }\textbf {\bibinfo {volume}
  {112}},\ \bibinfo {pages} {030602} (\bibinfo {year} {2014})}\BibitemShut
  {NoStop}%
\bibitem [{\citenamefont {Chand}\ and\ \citenamefont
  {Biswas}(2017)}]{chand2017measurement}%
  \BibitemOpen
  \bibfield  {author} {\bibinfo {author} {\bibfnamefont {S.}~\bibnamefont
  {Chand}}\ and\ \bibinfo {author} {\bibfnamefont {A.}~\bibnamefont {Biswas}},\
  }\href {https://doi.org/10.1103/PhysRevE.95.032111} {\bibfield  {journal}
  {\bibinfo  {journal} {Physical Review E}\ }\textbf {\bibinfo {volume} {95}},\
  \bibinfo {pages} {032111} (\bibinfo {year} {2017})}\BibitemShut {NoStop}%
\bibitem [{\citenamefont {Klaers}\ \emph {et~al.}(2017)\citenamefont {Klaers},
  \citenamefont {Faelt}, \citenamefont {Imamoglu},\ and\ \citenamefont
  {Togan}}]{klaers2017squeezed}%
  \BibitemOpen
  \bibfield  {author} {\bibinfo {author} {\bibfnamefont {J.}~\bibnamefont
  {Klaers}}, \bibinfo {author} {\bibfnamefont {S.}~\bibnamefont {Faelt}},
  \bibinfo {author} {\bibfnamefont {A.}~\bibnamefont {Imamoglu}},\ and\
  \bibinfo {author} {\bibfnamefont {E.}~\bibnamefont {Togan}},\ }\href
  {https://doi.org/10.1103/PhysRevX.7.031044} {\bibfield  {journal} {\bibinfo
  {journal} {Physical Review X}\ }\textbf {\bibinfo {volume} {7}},\ \bibinfo
  {pages} {031044} (\bibinfo {year} {2017})}\BibitemShut {NoStop}%
\bibitem [{\citenamefont {Camati}\ \emph {et~al.}(2019)\citenamefont {Camati},
  \citenamefont {Santos},\ and\ \citenamefont {Serra}}]{camati2019coherence}%
  \BibitemOpen
  \bibfield  {author} {\bibinfo {author} {\bibfnamefont {P.~A.}\ \bibnamefont
  {Camati}}, \bibinfo {author} {\bibfnamefont {J.~F.}\ \bibnamefont {Santos}},\
  and\ \bibinfo {author} {\bibfnamefont {R.~M.}\ \bibnamefont {Serra}},\ }\href
  {https://doi.org/10.1103/PhysRevA.99.062103} {\bibfield  {journal} {\bibinfo
  {journal} {Physical Review A}\ }\textbf {\bibinfo {volume} {99}},\ \bibinfo
  {pages} {062103} (\bibinfo {year} {2019})}\BibitemShut {NoStop}%
\bibitem [{\citenamefont {Ro{\ss}nagel}\ \emph {et~al.}(2016)\citenamefont
  {Ro{\ss}nagel}, \citenamefont {Dawkins}, \citenamefont {Tolazzi},
  \citenamefont {Abah}, \citenamefont {Lutz}, \citenamefont {Schmidt-Kaler},\
  and\ \citenamefont {Singer}}]{rossnagel2016single}%
  \BibitemOpen
  \bibfield  {author} {\bibinfo {author} {\bibfnamefont {J.}~\bibnamefont
  {Ro{\ss}nagel}}, \bibinfo {author} {\bibfnamefont {S.~T.}\ \bibnamefont
  {Dawkins}}, \bibinfo {author} {\bibfnamefont {K.~N.}\ \bibnamefont
  {Tolazzi}}, \bibinfo {author} {\bibfnamefont {O.}~\bibnamefont {Abah}},
  \bibinfo {author} {\bibfnamefont {E.}~\bibnamefont {Lutz}}, \bibinfo {author}
  {\bibfnamefont {F.}~\bibnamefont {Schmidt-Kaler}},\ and\ \bibinfo {author}
  {\bibfnamefont {K.}~\bibnamefont {Singer}},\ }\href
  {https://www.science.org/doi/10.1126/science.aad6320} {\bibfield  {journal}
  {\bibinfo  {journal} {Science}\ }\textbf {\bibinfo {volume} {352}},\ \bibinfo
  {pages} {325} (\bibinfo {year} {2016})}\BibitemShut {NoStop}%
\bibitem [{\citenamefont {Levy}\ and\ \citenamefont
  {Kosloff}(2012)}]{levy2012quantum}%
  \BibitemOpen
  \bibfield  {author} {\bibinfo {author} {\bibfnamefont {A.}~\bibnamefont
  {Levy}}\ and\ \bibinfo {author} {\bibfnamefont {R.}~\bibnamefont {Kosloff}},\
  }\href {https://doi.org/https://doi.org/10.1103/PhysRevLett.108.070604}
  {\bibfield  {journal} {\bibinfo  {journal} {Physical review letters}\
  }\textbf {\bibinfo {volume} {108}},\ \bibinfo {pages} {070604} (\bibinfo
  {year} {2012})}\BibitemShut {NoStop}%
\bibitem [{\citenamefont {Palao}\ \emph {et~al.}(2001)\citenamefont {Palao},
  \citenamefont {Kosloff},\ and\ \citenamefont {Gordon}}]{palao2001quantum}%
  \BibitemOpen
  \bibfield  {author} {\bibinfo {author} {\bibfnamefont {J.~P.}\ \bibnamefont
  {Palao}}, \bibinfo {author} {\bibfnamefont {R.}~\bibnamefont {Kosloff}},\
  and\ \bibinfo {author} {\bibfnamefont {J.~M.}\ \bibnamefont {Gordon}},\
  }\href {https://doi.org/https://doi.org/10.1103/PhysRevE.64.056130}
  {\bibfield  {journal} {\bibinfo  {journal} {Physical Review E}\ }\textbf
  {\bibinfo {volume} {64}},\ \bibinfo {pages} {056130} (\bibinfo {year}
  {2001})}\BibitemShut {NoStop}%
\bibitem [{\citenamefont {Linden}\ \emph {et~al.}(2010)\citenamefont {Linden},
  \citenamefont {Popescu},\ and\ \citenamefont {Skrzypczyk}}]{linden2010small}%
  \BibitemOpen
  \bibfield  {author} {\bibinfo {author} {\bibfnamefont {N.}~\bibnamefont
  {Linden}}, \bibinfo {author} {\bibfnamefont {S.}~\bibnamefont {Popescu}},\
  and\ \bibinfo {author} {\bibfnamefont {P.}~\bibnamefont {Skrzypczyk}},\
  }\href {https://doi.org/https://doi.org/10.1103/PhysRevLett.105.130401}
  {\bibfield  {journal} {\bibinfo  {journal} {Physical review letters}\
  }\textbf {\bibinfo {volume} {105}},\ \bibinfo {pages} {130401} (\bibinfo
  {year} {2010})}\BibitemShut {NoStop}%
\bibitem [{\citenamefont {Hofer}\ \emph {et~al.}()\citenamefont {Hofer},
  \citenamefont {Perarnau-Llobet}, \citenamefont {Brask}, \citenamefont
  {Silva}, \citenamefont {Huber},\ and\ \citenamefont
  {Brunner}}]{hofer_autonomous_2016}%
  \BibitemOpen
  \bibfield  {author} {\bibinfo {author} {\bibfnamefont {P.~P.}\ \bibnamefont
  {Hofer}}, \bibinfo {author} {\bibfnamefont {M.}~\bibnamefont
  {Perarnau-Llobet}}, \bibinfo {author} {\bibfnamefont {J.~B.}\ \bibnamefont
  {Brask}}, \bibinfo {author} {\bibfnamefont {R.}~\bibnamefont {Silva}},
  \bibinfo {author} {\bibfnamefont {M.}~\bibnamefont {Huber}},\ and\ \bibinfo
  {author} {\bibfnamefont {N.}~\bibnamefont {Brunner}},\ }\href
  {https://doi.org/10.1103/PhysRevB.94.235420} {\bibfield  {journal} {\bibinfo
  {journal} {Physical Review B}\ }\textbf {\bibinfo {volume} {94}},\ \bibinfo
  {pages} {235420}}\BibitemShut {NoStop}%
\bibitem [{\citenamefont {Maslennikov}\ \emph {et~al.}()\citenamefont
  {Maslennikov}, \citenamefont {Ding}, \citenamefont {Hablützel},
  \citenamefont {Gan}, \citenamefont {Roulet}, \citenamefont {Nimmrichter},
  \citenamefont {Dai}, \citenamefont {Scarani},\ and\ \citenamefont
  {Matsukevich}}]{maslennikov_quantum_2019}%
  \BibitemOpen
  \bibfield  {author} {\bibinfo {author} {\bibfnamefont {G.}~\bibnamefont
  {Maslennikov}}, \bibinfo {author} {\bibfnamefont {S.}~\bibnamefont {Ding}},
  \bibinfo {author} {\bibfnamefont {R.}~\bibnamefont {Hablützel}}, \bibinfo
  {author} {\bibfnamefont {J.}~\bibnamefont {Gan}}, \bibinfo {author}
  {\bibfnamefont {A.}~\bibnamefont {Roulet}}, \bibinfo {author} {\bibfnamefont
  {S.}~\bibnamefont {Nimmrichter}}, \bibinfo {author} {\bibfnamefont
  {J.}~\bibnamefont {Dai}}, \bibinfo {author} {\bibfnamefont {V.}~\bibnamefont
  {Scarani}},\ and\ \bibinfo {author} {\bibfnamefont {D.}~\bibnamefont
  {Matsukevich}},\ }\href {https://doi.org/10.1038/s41467-018-08090-0}
  {\bibfield  {journal} {\bibinfo  {journal} {Nature Communications}\ }\textbf
  {\bibinfo {volume} {10}},\ \bibinfo {pages} {202}},\ \bibinfo {note} {number:
  1 Publisher: Nature Publishing Group}\BibitemShut {NoStop}%
\bibitem [{\citenamefont {Nimmrichter}\ \emph {et~al.}()\citenamefont
  {Nimmrichter}, \citenamefont {Dai}, \citenamefont {Roulet},\ and\
  \citenamefont {Scarani}}]{nimmrichter_quantum_2017}%
  \BibitemOpen
  \bibfield  {author} {\bibinfo {author} {\bibfnamefont {S.}~\bibnamefont
  {Nimmrichter}}, \bibinfo {author} {\bibfnamefont {J.}~\bibnamefont {Dai}},
  \bibinfo {author} {\bibfnamefont {A.}~\bibnamefont {Roulet}},\ and\ \bibinfo
  {author} {\bibfnamefont {V.}~\bibnamefont {Scarani}},\ }\href
  {https://doi.org/10.22331/q-2017-12-11-37} {\bibfield  {journal} {\bibinfo
  {journal} {Quantum}\ }\textbf {\bibinfo {volume} {1}},\ \bibinfo {pages}
  {37}},\ \Eprint {https://arxiv.org/abs/1709.08353 [quant-ph]} {1709.08353
  [quant-ph]} \BibitemShut {NoStop}%
\bibitem [{\citenamefont {Dodonov}(2020)}]{dodonov2020fifty}%
  \BibitemOpen
  \bibfield  {author} {\bibinfo {author} {\bibfnamefont {V.}~\bibnamefont
  {Dodonov}},\ }\href@noop {} {\bibfield  {journal} {\bibinfo  {journal}
  {Physics}\ }\textbf {\bibinfo {volume} {2}},\ \bibinfo {pages} {67} (\bibinfo
  {year} {2020})}\BibitemShut {NoStop}%
\bibitem [{\citenamefont {Lupascu}\ \emph {et~al.}()\citenamefont {Lupascu},
  \citenamefont {Harmans},\ and\ \citenamefont {Mooij}}]{lupascu_quantum_2005}%
  \BibitemOpen
  \bibfield  {author} {\bibinfo {author} {\bibfnamefont {A.}~\bibnamefont
  {Lupascu}}, \bibinfo {author} {\bibfnamefont {C.~J. P.~M.}\ \bibnamefont
  {Harmans}},\ and\ \bibinfo {author} {\bibfnamefont {J.~E.}\ \bibnamefont
  {Mooij}},\ }\href {https://doi.org/10.1103/PhysRevB.71.184506} {\bibfield
  {journal} {\bibinfo  {journal} {Physical Review B}\ }\textbf {\bibinfo
  {volume} {71}},\ \bibinfo {pages} {184506}},\ \Eprint
  {https://arxiv.org/abs/cond-mat/0410730} {cond-mat/0410730} \BibitemShut
  {NoStop}%
\bibitem [{cla()}]{clarke_squid_2004}%
  \BibitemOpen
  \href@noop {} {\bibinfo {title} {The {SQUID} handbook}},\ \bibinfo {note}
  {{OCLC}: ocm52746892}\BibitemShut {NoStop}%
\bibitem [{\citenamefont {Wilson}\ \emph {et~al.}()\citenamefont {Wilson},
  \citenamefont {Johansson}, \citenamefont {Pourkabirian}, \citenamefont
  {Simoen}, \citenamefont {Johansson}, \citenamefont {Duty}, \citenamefont
  {Nori},\ and\ \citenamefont {Delsing}}]{wilson_observation_2011}%
  \BibitemOpen
  \bibfield  {author} {\bibinfo {author} {\bibfnamefont {C.~M.}\ \bibnamefont
  {Wilson}}, \bibinfo {author} {\bibfnamefont {G.}~\bibnamefont {Johansson}},
  \bibinfo {author} {\bibfnamefont {A.}~\bibnamefont {Pourkabirian}}, \bibinfo
  {author} {\bibfnamefont {M.}~\bibnamefont {Simoen}}, \bibinfo {author}
  {\bibfnamefont {J.~R.}\ \bibnamefont {Johansson}}, \bibinfo {author}
  {\bibfnamefont {T.}~\bibnamefont {Duty}}, \bibinfo {author} {\bibfnamefont
  {F.}~\bibnamefont {Nori}},\ and\ \bibinfo {author} {\bibfnamefont
  {P.}~\bibnamefont {Delsing}},\ }\href {https://doi.org/10.1038/nature10561}
  {\bibfield  {journal} {\bibinfo  {journal} {Nature}\ }\textbf {\bibinfo
  {volume} {479}},\ \bibinfo {pages} {376}}\BibitemShut {NoStop}%
\bibitem [{\citenamefont {Felicetti}\ \emph {et~al.}()\citenamefont
  {Felicetti}, \citenamefont {Sanz}, \citenamefont {Lamata}, \citenamefont
  {Romero}, \citenamefont {Johansson}, \citenamefont {Delsing},\ and\
  \citenamefont {Solano}}]{felicetti_dynamical_2014}%
  \BibitemOpen
  \bibfield  {author} {\bibinfo {author} {\bibfnamefont {S.}~\bibnamefont
  {Felicetti}}, \bibinfo {author} {\bibfnamefont {M.}~\bibnamefont {Sanz}},
  \bibinfo {author} {\bibfnamefont {L.}~\bibnamefont {Lamata}}, \bibinfo
  {author} {\bibfnamefont {G.}~\bibnamefont {Romero}}, \bibinfo {author}
  {\bibfnamefont {G.}~\bibnamefont {Johansson}}, \bibinfo {author}
  {\bibfnamefont {P.}~\bibnamefont {Delsing}},\ and\ \bibinfo {author}
  {\bibfnamefont {E.}~\bibnamefont {Solano}},\ }\href
  {https://doi.org/10.1103/PhysRevLett.113.093602} {\bibfield  {journal}
  {\bibinfo  {journal} {Physical Review Letters}\ }\textbf {\bibinfo {volume}
  {113}},\ \bibinfo {pages} {093602}},\ \Eprint
  {https://arxiv.org/abs/1402.4451 [cond-mat, physics:quant-ph]} {1402.4451
  [cond-mat, physics:quant-ph]} \BibitemShut {NoStop}%
\bibitem [{\citenamefont {Wallquist}\ \emph {et~al.}()\citenamefont
  {Wallquist}, \citenamefont {Shumeiko},\ and\ \citenamefont
  {Wendin}}]{wallquist_selective_2006}%
  \BibitemOpen
  \bibfield  {author} {\bibinfo {author} {\bibfnamefont {M.}~\bibnamefont
  {Wallquist}}, \bibinfo {author} {\bibfnamefont {V.~S.}\ \bibnamefont
  {Shumeiko}},\ and\ \bibinfo {author} {\bibfnamefont {G.}~\bibnamefont
  {Wendin}},\ }\href {https://doi.org/10.1103/PhysRevB.74.224506} {\bibfield
  {journal} {\bibinfo  {journal} {Physical Review B}\ }\textbf {\bibinfo
  {volume} {74}},\ \bibinfo {pages} {224506}},\ \Eprint
  {https://arxiv.org/abs/cond-mat/0608209} {cond-mat/0608209} \BibitemShut
  {NoStop}%
\bibitem [{\citenamefont {Bourassa}\ \emph {et~al.}()\citenamefont {Bourassa},
  \citenamefont {Beaudoin}, \citenamefont {Gambetta},\ and\ \citenamefont
  {Blais}}]{bourassa_josephson_2012}%
  \BibitemOpen
  \bibfield  {author} {\bibinfo {author} {\bibfnamefont {J.}~\bibnamefont
  {Bourassa}}, \bibinfo {author} {\bibfnamefont {F.}~\bibnamefont {Beaudoin}},
  \bibinfo {author} {\bibfnamefont {J.~M.}\ \bibnamefont {Gambetta}},\ and\
  \bibinfo {author} {\bibfnamefont {A.}~\bibnamefont {Blais}},\ }\href
  {https://doi.org/10.1103/PhysRevA.86.013814} {\bibfield  {journal} {\bibinfo
  {journal} {Physical Review A}\ }\textbf {\bibinfo {volume} {86}},\ \bibinfo
  {pages} {013814}},\ \Eprint {https://arxiv.org/abs/1204.2237 [cond-mat,
  physics:quant-ph]} {1204.2237 [cond-mat, physics:quant-ph]} \BibitemShut
  {NoStop}%
\bibitem [{\citenamefont {Law}()}]{law_interaction_1995}%
  \BibitemOpen
  \bibfield  {author} {\bibinfo {author} {\bibfnamefont {C.~K.}\ \bibnamefont
  {Law}},\ }\href {https://doi.org/10.1103/PhysRevA.51.2537} {\bibfield
  {journal} {\bibinfo  {journal} {Physical Review A}\ }\textbf {\bibinfo
  {volume} {51}},\ \bibinfo {pages} {2537}}\BibitemShut {NoStop}%
\bibitem [{\citenamefont {Velasco}\ \emph {et~al.}(2022)\citenamefont
  {Velasco}, \citenamefont {Del~Grosso}, \citenamefont {Lombardo},
  \citenamefont {Soba},\ and\ \citenamefont {Villar}}]{velasco2022photon}%
  \BibitemOpen
  \bibfield  {author} {\bibinfo {author} {\bibfnamefont {C.~I.}\ \bibnamefont
  {Velasco}}, \bibinfo {author} {\bibfnamefont {N.~F.}\ \bibnamefont
  {Del~Grosso}}, \bibinfo {author} {\bibfnamefont {F.~C.}\ \bibnamefont
  {Lombardo}}, \bibinfo {author} {\bibfnamefont {A.}~\bibnamefont {Soba}},\
  and\ \bibinfo {author} {\bibfnamefont {P.~I.}\ \bibnamefont {Villar}},\
  }\href {https://doi.org/https://doi.org/10.1103/PhysRevA.106.043701}
  {\bibfield  {journal} {\bibinfo  {journal} {Physical Review A}\ }\textbf
  {\bibinfo {volume} {106}},\ \bibinfo {pages} {043701} (\bibinfo {year}
  {2022})}\BibitemShut {NoStop}%
\bibitem [{\citenamefont {Drechsler}\ \emph {et~al.}(2020)\citenamefont
  {Drechsler}, \citenamefont {Far{\'\i}as}, \citenamefont {Freitas},
  \citenamefont {Schmiegelow},\ and\ \citenamefont {Paz}}]{drechsler2020state}%
  \BibitemOpen
  \bibfield  {author} {\bibinfo {author} {\bibfnamefont {M.}~\bibnamefont
  {Drechsler}}, \bibinfo {author} {\bibfnamefont {M.~B.}\ \bibnamefont
  {Far{\'\i}as}}, \bibinfo {author} {\bibfnamefont {N.}~\bibnamefont
  {Freitas}}, \bibinfo {author} {\bibfnamefont {C.~T.}\ \bibnamefont
  {Schmiegelow}},\ and\ \bibinfo {author} {\bibfnamefont {J.~P.}\ \bibnamefont
  {Paz}},\ }\href {https://doi.org/https://doi.org/10.1103/PhysRevA.101.052331}
  {\bibfield  {journal} {\bibinfo  {journal} {Physical Review A}\ }\textbf
  {\bibinfo {volume} {101}},\ \bibinfo {pages} {052331} (\bibinfo {year}
  {2020})}\BibitemShut {NoStop}%
\bibitem [{\citenamefont {Johansson}\ \emph {et~al.}(2012)\citenamefont
  {Johansson}, \citenamefont {Nation},\ and\ \citenamefont
  {Nori}}]{johansson2012qutip}%
  \BibitemOpen
  \bibfield  {author} {\bibinfo {author} {\bibfnamefont {J.~R.}\ \bibnamefont
  {Johansson}}, \bibinfo {author} {\bibfnamefont {P.~D.}\ \bibnamefont
  {Nation}},\ and\ \bibinfo {author} {\bibfnamefont {F.}~\bibnamefont {Nori}},\
  }\href {https://doi.org/https://doi.org/10.1016/j.cpc.2012.11.019} {\bibfield
   {journal} {\bibinfo  {journal} {Computer Physics Communications}\ }\textbf
  {\bibinfo {volume} {183}},\ \bibinfo {pages} {1760} (\bibinfo {year}
  {2012})}\BibitemShut {NoStop}%
\bibitem [{\citenamefont {Freitas}\ and\ \citenamefont
  {Paz}(2017)}]{freitas2017fundamental}%
  \BibitemOpen
  \bibfield  {author} {\bibinfo {author} {\bibfnamefont {N.}~\bibnamefont
  {Freitas}}\ and\ \bibinfo {author} {\bibfnamefont {J.~P.}\ \bibnamefont
  {Paz}},\ }\href {https://doi.org/https://doi.org/10.1103/PhysRevE.95.012146}
  {\bibfield  {journal} {\bibinfo  {journal} {Physical Review E}\ }\textbf
  {\bibinfo {volume} {95}},\ \bibinfo {pages} {012146} (\bibinfo {year}
  {2017})}\BibitemShut {NoStop}%
\end{thebibliography}%
\pagebreak
\onecolumngrid
\appendix 
\section{Lagrangian of the cavity modes}\label{app:lagrangian}
To obtain the Lagrangian of the cavity $L_{cav}$ we first need to find 
\begin{gather}
  \dot{\Phi}=\sum_{n}\dot{\phi}_{n}\psi_{n}+\phi_{n}\dot{\psi}\\  
  \Phi'(x,t)=\sum_{n}\phi_{n}(t)\psi'_{n}(x,t)
\end{gather}
and consequently 
\begin{gather}
\dot{\Phi}^{2}=\sum_{n,m}\dot{\phi}_{n}\dot{\phi}_{m}\psi_{n}\psi_{m}+\sum_{n,m}\phi_{n}\phi_{m}\dot{\psi}_{n}\dot{\psi}_{m}+2\sum_{n,m}\dot{\phi}_{n}\phi_{m}\psi_{n}\dot{\psi}_{m}\\
\Phi'^2(x,t)=\sum_{mn}\phi_{n}(t)\phi_{m}(t)\psi'_{n}(x,t)\psi'_{m}(x,t).
\end{gather}
We then replace these terms inside the Lagrangian \eqref{eq:lagrangian_cav} and from the first term we will have 
\begin{gather}
   \left(\frac{\hbar}{2e}\right)^2 \mathcal{C}_0\int_{-d/2}^{d/2}(1+2C_{J}/\mathcal{C}_0\delta(x))\dot{\Phi}^{2}= \left(\frac{\hbar}{2e}\right)^2\mathcal{C}_0\sum_{n,m}\dot{\phi}_{n}\dot{\phi}_{m}\int_{-d/2}^{d/2}(1+2C_{J}/\mathcal{C}_0\delta(x))\psi_{n}\psi_{m}\notag\\
   +\left(\frac{\hbar}{2e}\right)^2\mathcal{C}_0\sum_{n,m}\phi_{n}\phi_{m}\int_{-d/2}^{d/2}(1+2C_{J}/\mathcal{C}_0\delta(x))\dot{\psi}_{n}\dot{\psi}_{m}+2\left(\frac{\hbar}{2e}\right)^2\mathcal{C}_0\sum_{n,m}\dot{\phi}_{n}\phi_{m}\int_{-d/2}^{d/2}(1+2C_{J}/\mathcal{C}_0\delta(x))\psi_{n}\dot{\psi}_{m}.
\end{gather}
We define the inner product as 
\begin{gather}
    \frac{1}{d}\int_{-d/2}^{d/2}(1+2C_{J}/\mathcal{C}_0\delta(x))\psi_{n}\psi_{m}=\delta_{mn},
\end{gather}
and using this we obtain 
\begin{equation}
   \left(\frac{\hbar}{2e}\right)^2 \mathcal{C}_0\int_{-d/2}^{d/2}(1+2C_{J}/\mathcal{C}_0\delta(x))\dot{\Phi}^{2}=\left(\frac{\hbar}{2e}\right)^2C\sum_{n}\dot{\phi}_{n}^{2}+2\left(\frac{\hbar}{2e}\right)^2C\sum_{n,m}A_{mn}\dot{\phi}_{n}\phi_{m}+\left(\frac{\hbar}{2e}\right)^2C\sum_{n,m}B_{mn}\phi_{n}\phi_{m}
\end{equation}
where we defined $\mathcal{C}_0=C/d$ and 
\begin{gather}
 A_{mn}=\frac{1}{d}\int_{-d/2}^{d/2}dx(1+2C_{J}/\mathcal{C}_0\delta(x))\psi_{n}\dot{\psi}_{m} \\
   B_{mn}=\frac{1}{d}\int_{-d/2}^{d/2}dx(1+2C_{J}/\mathcal{C}_0\delta(x))\dot{\psi}_{n}\dot{\psi}_{m}=\sum_kA_{mk}A_{nk}
=\sum_kA_{mk}A_{nk}.
\end{gather}
The next term in the Lagrangian will give rise to 
\begin{gather}
    -\int_{-d/2}^{d/2}dxv^{2}\Phi'^{2}=-v^{2}\sum_{mn}\phi_{n}(t)\phi_{m}(t)\int_{-d/2}^{d/2}dx\psi'_{n}(x,t)\psi'_{m}(x,t)=-v^{2}\sum_{mn}k_{n}^{2}\phi_{n}^{2}(t)+E_{J}\cos f\Phi_{0},
\end{gather}
where to obtain the last equality we have used the boundary condition \eqref{eq:boundary_deriv} together with the inner product. Therefore the Lagrangian will be written as 
\section{Hamiltonian of the system}\label{app:Hamiltonian}
After doing the Legendre transformation of the Lagrangian in Eq.~\eqref{eq:lagrang_total} we will find the Hamiltonian such that 
\begin{gather}
H=\frac{1}{2}\left(\frac{\hbar}{2e}\right)^{2}C\sum_{n}\left(\dot{\phi_{n}}^{2}+\omega_{n}^{2}\phi_{n}^{2}\right)+\hbar\left(\frac{1}{2e}\right)^{2}C\dot{f}\sum_{n,m}M_{nm}\dot{\phi_{n}}\phi_{m}\notag\\
+\left(\frac{\hbar}{2e}\right)^{2}C_{J}\frac{\dot{f}^{2}}{2}+\left(\frac{\hbar}{2e}\right)^{2}C\frac{\dot{f}^{2}}{2}\sum_{n,m,k}M_{nk}M_{mk}\phi_{n}\phi_{m}+V(f).\label{eq:hamilt_full_f}
\end{gather}
Now we need to replace the momentum variables above. We notice that 
\begin{gather}
   \dot{\phi}_{n}=\frac{1}{\hbar\left(\frac{1}{2e}\right)^{2}C}q_{n}-\dot{f}M_{nm}\phi_{m},
   \end{gather}
   thus, using this we can write 
   \begin{equation}
       p_{f}=\sum_{n,m}M_{nm}q_{n}\phi_{m}+\hbar\left(\frac{1}{2e}\right)^{2}C_{J}\dot{f},
   \end{equation}
   and hence 
   \begin{equation}
       \dot{f}=\frac{1}{\hbar\left(\frac{1}{2e}\right)^{2}C_{J}}p_{f}-\frac{1}{\hbar\left(\frac{1}{2e}\right)^{2}C_{J}}\sum_{n,m}M_{nm}q_{n}\phi_{m}.
   \end{equation}
   Replacing  $\dot{\phi}_n$ in the Hamiltonian \eqref{eq:hamilt_full_f} we find
   \begin{gather}
      H=\sum_{n}\left[\frac{\left(2e\right)^{2}}{2C}q_{n}^{2}+\frac{1}{2}\left(\frac{\hbar}{2e}\right)^{2}C\omega_{n}^{2}\phi_{n}^{2}\right]+\left(\frac{\hbar}{2e}\right)^{2}C_{J}\frac{\dot{ f}^{2}}{2}+V(f).
   \end{gather}
   Therefore the final Hamiltonian after replacing $\dot{\delta f}$ as well will take the following form 
   \begin{equation}
     H=\sum_{n}\left[\frac{\left(2e\right)^{2}}{2C}q_{n}^{2}+\frac{1}{2}\left(\frac{\hbar}{2e}\right)^{2}C\omega_{n}^{2}\phi_{n}^{2}\right]+\frac{\left(2e\right)^{2}}{2C_{J}}\left(p_{f}-\sum_{n,m}M_{nm}q_{n}\phi_{m}\right)^{2}+V(f).  
   \end{equation}
   To go further we first quantize the full Hamiltonian by assuming the usual commutation reation among the operators. Therefore we write 
 \begin{gather}
       \phi_{n}=\sqrt{\frac{\hbar\omega_{n}}{2E_{C}}}\left(a_{n}+a_{n}^{\dagger}\right)\\
       q_{n}=-i\sqrt{\frac{E_{C}}{2\hbar\omega_{n}}}\left(a_{n}-a_{n}^{\dagger}\right).
   \end{gather}
Thus the Hamiltonian will become 
\begin{gather}\label{eq:hamil_exactA}
     H=\sum_{n}\hbar\omega_{n}a_{n}^{\dagger}a_{n}+\frac{\left(2e\right)^{2}}{2C_{J}}\left(p_{f}+\Gamma(f)\right)^{2}+V(f), 
\end{gather}
where $\Gamma (f)=\frac{i}{2}\sum_{n,m}M_{nm}\left(a_{n}-a_{n}^{\dagger}\right)\left(a_{n}+a_{n}^{\dagger}\right)$. We notice that the operators $a_n$ and $a^{\dagger}_n$ depend on $\delta f$ through $\omega_{n}$. Therefore, there exists an interaction among the cavity fields and the SQUID degrees of freedom $f$. This said, we can now do the first linear approximation of the Hamiltonian. This can be done by assuming that the position of the SQUID, $f$, has small oscillations around its rest position, $f_0$.

Explicitly, we can write for $f(t)\approx f_{0}+\delta f(t)$ where $\delta f(t)\ll1$;
\begin{equation}
    \omega_{n}(f)\approx \omega_{n}(f_{0})+\delta f\omega_{n}^{\prime}(f_{0}).
\end{equation}
Therefore 
\begin{equation}
   a_{n}\approx \sqrt{\frac{E_{C}}{2\hbar\omega_{n}(f_{0})}}\phi_{n}+i\sqrt{\frac{\hbar\omega_{n}(f_{0})}{2E_{C}}}q_{n}-\frac{1}{2}\delta f\frac{\omega_{n}^{\prime}(f_{0})}{\omega_{n}(f_{0})}\left(\sqrt{\frac{E_{C}}{2\hbar\omega_{n}(f_{0})}}\phi_{n}-i\sqrt{\frac{\hbar\omega_{n}(f_{0})}{2E_{C}}}q_{n}\right)
\end{equation}
which can be written as 
\begin{equation}\label{eq:anni_approx}
    a_{n}\approx a_{n}(f_{0})-\frac{1}{2}\delta f\frac{\omega_{n}^{\prime}(f_{0})}{\omega_{n}(f_{0})}a_{n}^{\dagger}(f_{0}).
\end{equation}
Moreover, we notice that in this limit using Eq.~\eqref{eq:M} we have 
\begin{equation}
    M_{nm}=M_{nm0}+\delta f M'_{nm0},
\end{equation}
with $M'_{mn0}=\sum_{k}M_{nk0}M_{mk0}$ and 
\begin{equation}
   M_{nm0} =\frac{1}{d}\int_{-d/2}^{d/2}dx(1+2C_{J}/\mathcal{C}_0\delta(x))\psi_{m}(f_{0})\frac{dk_{n}}{df_{0}}\frac{d\psi_{n}(f_{0})}{dk_{n}}.
\end{equation}
Now we should replace all these terms back into the bare Hamiltonian and $\Gamma (f)$. In this way,
these terms give nonlinear contributions to the interaction Hamiltonian. Therefore, using the fact that $\delta f\ll 1$, we can do a linear approximation. To do so we write $\Gamma(f)\approx\Gamma(f_0)+\delta f \Gamma'(f_0)$ where $\Gamma'$ can be found by replacing the above expansions of creation and annihilation operators and $M_nm$ in $\Gamma (f)$ and only keeping the first order in $\delta f$;
\begin{equation}
    \Gamma'(f_0)=\sqrt{\frac{\omega_{m}(f_{0})}{\omega_{n}(f_{0})}}\left[M_{nm}^{\prime}+M_{nm}\frac{\omega_{n}^{\prime}(f_{0})}{\omega_{n}(f_{0})}\right]\left(a_{m}(f_{0})+a_{m}^{\dagger}(f_{0})\right)\left(a_{n}(f_{0})-a_{n}^{\dagger}(f_{0})\right).
\end{equation}
Replacing this into the Hamiltonian we will have 
\begin{gather}
    H=\hbar\sum_{n}\left(\omega_{n}+\omega_{n}^{\prime}\delta f\right)\left(a_{n}^{\dagger}-\delta f\frac{\omega_{n}^{\prime}}{2\omega_{n}}a_{n}\right)\left(a_{n}-\delta f\frac{\omega_{n}^{\prime}}{2\omega_{n}}a_{n}^{\dagger}\right)+\frac{\left(2e\right)^{2}}{2C_{J}}\left(p_{f}+\Gamma_0+\delta f\Gamma'_0\right)^{2}+V(\delta f),
\end{gather}
where we just for simplicity replaced $\Gamma(f_0)$ with $\Gamma_0$ and the same for $\Gamma'$. Moreover
\begin{equation}
    V(\delta f)=\left(\frac{\hbar}{2e}\right)^{2}\frac{1}{2}C_{J}\omega_{f}^{2}\delta f^{2}+\left(\frac{\hbar}{2e}\right)^{2}\frac{2M}{LL_{\rm ext}}\delta ff_{\rm ext}.
\end{equation}
In the above Hamiltonian, the ladder operators are independent of $f$. To find the form of the interaction, we perform a unitary transformation $H'=T^{\dagger} H T$ where 
\begin{equation}
    T=\exp\left\{ i\delta f(\Gamma_0+\frac{1}{2}\delta f\Gamma'_0)\right\}. 
\end{equation}
We first apply this on the second term of the Hamiltonain and the result will be 
\begin{equation}
    T^{\dagger} \left(p_{f}+\Gamma_0+\delta f\Gamma'_0\right)^{2} T=p_f^2,
\end{equation}
so we are shifting the momentum by $\Gamma_0+\delta f\Gamma'_0$. Next we need to transform the first term in the Hamiltonian,
\begin{gather}
   T^{\dagger} \hbar\sum_{n}\left(\omega_{n}+\omega_{n}^{\prime}\delta f\right)\left(a_{n}^{\dagger}-\delta f\frac{\omega_{n}^{\prime}}{2\omega_{n}}a_{n}\right)\left(a_{n}-\delta f\frac{\omega_{n}^{\prime}}{2\omega_{n}}a_{n}^{\dagger}\right)T\approx\notag\\
   \hbar\sum_{n}\omega_{n}\left(a_{n}^{\dagger}a_{n}-\delta f\frac{\omega_{n}^{\prime}}{2\omega_{n}}(a_{n}^{\dagger2}+a_{n}^{2})+\delta f^{2}\left(\frac{\omega_{n}^{\prime}}{2\omega_{n}}\right)^{2}a_{n}a_{n}^{\dagger}\right)+\hbar\sum_{n}\omega_{n}^{\prime}\left(\delta fa_{n}^{\dagger}a_{n}-\delta f^{2}\frac{\omega_{n}^{\prime}}{2\omega_{n}}(a_{n}^{\dagger2}+a_{n}^{2})\right)\notag\\
    \frac{-\hbar}{2}\delta f\sum_{n,m}M_{nm}\sqrt{\frac{\omega_{m}}{\omega_{n}}}\left[\omega_{m}\left(a_{m}-a_{m}^{\dagger}\right)\left(a_{n}-a_{n}^{\dagger}\right)+\omega_{n}\left(a_{m}+a_{m}^{\dagger}\right)\left(a_{n}+a_{n}^{\dagger}\right)\right]\notag\\
    -\frac{-\hbar}{4}\delta f^{2}\sum_{n,m}\tilde{M}_{nm}\sqrt{\frac{\omega_{m}}{\omega_{n}}}\left[\omega_{m}\left(a_{m}-a_{m}^{\dagger}\right)\left(a_{n}-a_{n}^{\dagger}\right)+\omega_{n}\left(a_{m}+a_{m}^{\dagger}\right)\left(a_{n}+a_{n}^{\dagger}\right)\right]\notag\\
    -\frac{\hbar}{2}\delta f^{2}\sum_{mn}M_{nm}\sqrt{\frac{\omega_{m}}{\omega_{n}}}\left[\omega_{n}\frac{\omega_{n}^{\prime}}{\omega_{n}}\left(a_{m}+a_{m}^{\dagger}\right)\left(a_{n}^{\dagger}+a_{n}\right)+\omega_{m}\frac{\omega_{m}^{\prime}}{\omega_{m}}\left(a_{m}^{\dagger}-a_{m}\right)\left(a_{n}-a_{n}^{\dagger}\right)\right]\notag\\
    -\frac{\hbar}{2}\delta f^{2}\sum_{n,m}M_{nm}\sqrt{\frac{\omega_{m}}{\omega_{n}}}\left[\omega'_{m}\left(a_{m}-a_{m}^{\dagger}\right)\left(a_{n}-a_{n}^{\dagger}\right)+\omega'_{n}\left(a_{m}+a_{m}^{\dagger}\right)\left(a_{n}+a_{n}^{\dagger}\right)\right]+O(\delta f^3),\label{eq:interaction_deltaf}
\end{gather}
where $\tilde{M}_{nm}=M_{nm}^{\prime}+M_{nm}\frac{\omega_{n}^{\prime}}{\omega_{n}}$. Now these terms are denoting the interaction between the cavities and the squid in powers of $\delta f$ and we only kept terms up to the second order in $\delta f$. We can further simplify this Hamiltonian in two ways. First, since $\delta f$ is very small, we can neglect all second order terms in $\delta f$. Note that there exists a second order term in $V(\delta _f)$ which is written as $\left(\frac{\hbar}{2e}\right)^{2}\frac{1}{2}C_{J}\omega_{f}^{2}\delta f^{2}$. However, the prefactor of this term, $\left(\frac{\hbar}{2e}\right)^{2}\frac{1}{2}C_{J}=\hbar^2\omega_f^2/4E_{C_J}\propto E_J$ is large. The second simplification is the rotating wave approximation (RWA). One can show that all the terms of second order in $\delta f$ will be eliminated by the approximation. To sketch the RWA, we first write  
\begin{align}
       \delta f&=\sqrt{\frac{E_{C_{J}}}{\hbar\omega_{f}}}\left(a_{f}+a_{f}^{\dagger}\right)\\
       p_{f}&=-i\frac{1}{2}\sqrt{\frac{\hbar\omega_{f}}{E_{C_{J}}}}\left(a_{f}-a_{f}^{\dagger}\right).
         \end{align}
Replacing them in the Hamiltonian will result in a bare Hamiltonian of the form,
\begin{equation}
    H_0=\hbar\sum_n \omega_n a^{\dagger}_na_n+\hbar\omega_fa^{\dagger}_fa_f.
\end{equation}
Therefore we can use the above Hamiltonian and move to the rotating frame by applying the unitary transformation
\begin{equation}
    U=\exp \left\{ \frac{it}{\hbar}\left[\sum_{n}\hbar\omega_{n}a_{n}^{\dagger}a_{n}+\hbar\omega_{f}a_{f}^{\dagger}a_{f}\right] \right\}
\end{equation}
on the interaction Hamiltonian~\eqref{eq:interaction_deltaf}. Therefore, after doing the RWA, using the resonance condition $\omega_f=\omega_n+\omega_m$ for $m\neq n$, and going back to the Schr\"odinger picture, we obtain 
\begin{gather}
H_{\rm RWA}=\hbar\sum_{n}\omega_{n}a_{n}^{\dagger}a_{n}+\hbar\omega_{f}a_{f}^{\dagger}a_{f}+\hbar\sum_{n}\omega_{n}\frac{E_{C_{J}}}{\hbar\omega_{f}}\left(\frac{\omega_{n}^{\prime}}{2\omega_{n}}\right)^{2}a_{n}a_{n}^{\dagger}a_{f}a_{f}^{\dagger}\notag\\
  -\frac{\hbar}{2}\sqrt{\frac{E_{C_{J}}}{\hbar\omega_{f}}}\sum_{n,m}M_{nm}\sqrt{\frac{\omega_{m}}{\omega_{n}}}\left(\omega_{m}+\omega_{n}\right)\left(a_{f}^{\dagger}a_{m}a_{n}+a_{f}a_{m}^{\dagger}a_{n}^{\dagger}\right).
\end{gather}
Moreover, we can neglect the third term of the first line by reasoning that $E_{C_{J}}/\hbar\omega_{f}\propto\sqrt{E_{C_{J}}/ E_{J}}\ll 1$. Therefore that term is negligible and we can write the final Hamiltonian as 
\begin{gather}
H_{rwa}=\hbar\sum_{n}\omega_{n}a_{n}^{\dagger}a_{n}+\hbar\omega_{f}a_{f}^{\dagger}a_{f}
  -\frac{\hbar}{2}\sqrt{\frac{E_{C_{J}}}{\hbar\omega_{f}}}\sum_{n,m}M_{nm}\sqrt{\frac{\omega_{m}}{\omega_{n}}}\left(\omega_{m}+\omega_{n}\right)\left(a_{f}^{\dagger}a_{m}a_{n}+a_{f}a_{m}^{\dagger}a_{n}^{\dagger}\right).
\end{gather}

\section{the SQUID Lagrangian}

The phase drop of the SQUID over its inductance $L$ follows 
\begin{equation}
\left(\frac{\hbar}{2e}\right)^2C_{J}\ddot{f}+E_{J}\sin f+\left(\frac{\hbar}{2e}\right)^2\frac{2}{L}\left(f+\frac{M}{L_{\rm ext}}f_{\rm ext}\right)=0.\label{eq:eq_motion_f}
\end{equation}
Therefore its Lagrangian can be written as 
\begin{equation}
    \L=\left(\frac{\hbar}{2e}\right)^2\frac{C_{J}}{2}\dot{f}^2-V(f)
\end{equation}
where 
\begin{equation}
    V(f)=-E_{J}\cos f+\left(\frac{\hbar}{2e}\right)^2\frac{1}{L}\left(f^2+\frac{M}{L_{\rm ext}}ff_{\rm ext}\right).
\end{equation}
This Lagrangian is exact up to the assumption that $\Phi_0\ll 1$. However, we can also look at the limit of small displacement of $f$ around $f_0$ namely assuming 
\begin{gather}
  f=f_0+\delta f(t)\notag\\
  f_{\rm xt}=F_{\rm ext}+\delta f_{\rm ext}(t)
\end{gather}
where $\delta f\ll1$ and $\delta f_{\rm ext}\ll1$. This limit will be useful once we write the full Hamiltonian and we need to make the linear interaction approximation.

Therefore the equation of motion for $f$ can be rewritten as 
\begin{equation}
  \left(\frac{\hbar}{2e}\right)^2 C_{J}\ddot{\delta f}+\left(E_{J}\cos f_0+\left(\frac{\hbar}{2e}\right)^2\frac{2}{L}\right)\delta f+E_{J}\sin f_0+\left(\frac{\hbar}{2e}\right)^2\frac{2}{L}f_0=-\left(\frac{\hbar}{2e}\right)^2\frac{2M}{LL_{\rm ext}}\delta f_{\rm ext}-\left(\frac{\hbar}{2e}\right)^2\frac{2M}{LL_{\rm ext}}F_{\rm ext} 
\end{equation}
The stationary solution where we set $\delta f=0$ and $\delta f_{\rm ext}=0$ gives
\begin{equation}
  E_{J}\sin f_0+\left(\frac{\hbar}{2e}\right)^2\frac{2}{L}\left(f_0+\frac{M}{L_{\rm ext}}F_{\rm ext}\right)=0.  
\end{equation}
This indicates the relation between $f_0$ and $F_{\rm ext}$. Using it
we get 
\begin{equation}
  \left(\frac{\hbar}{2e}\right)^2C_{J}\ddot{\delta f}+\left(E_{J}\cos f_0+\left(\frac{\hbar}{2e}\right)^2\frac{2}{L}\right)\delta f=-\left(\frac{\hbar}{2e}\right)^2\frac{2M}{LL_{\rm ext}}\delta f_{\rm ext},  
\end{equation}
which is basically a forced oscillator. To quantize $f$, we first write the Hamiltonian using the above equation of motion. Since assuming $\Phi_0$ to be small will effectively decouple $f$
from the rest of the system, we can quantize its Hamiltonian independently.
One can rewrite the Lagrangian of $f$ according
to its equation of motion~\ref{eq:eq_motion_f}, which will lead to
\begin{equation}
   L_{f}=\left(\frac{\hbar}{2e}\right)^2C_{J}\frac{\dot{\delta f}^{2}}{2}-\left(\frac{\hbar}{2e}\right)^2\frac{1}{2}C_{J}\omega_{f}^{2}\delta f^{2}-\left(\frac{\hbar}{2e}\right)^2\frac{2M}{LL_{\rm ext}}\delta f\delta f_{\rm ext}
\end{equation}
where 
\begin{equation}
    \omega_{f}^{2}=\frac{1}{\left(\frac{\hbar}{2e}\right)^2C_{J}}\left(E_{J}\cos f_0+\left(\frac{\hbar}{2e}\right)^2\frac{2}{L}\right).
\end{equation}

\end{document}